\documentclass{nws}

\usepackage{latexsym, tabularx}
\usepackage{graphicx}
\usepackage{multicol, color, url}
\usepackage[authoryear]{natbib}
\usepackage[colorlinks,citecolor=blue,urlcolor=blue]{hyperref}

\def\btheta{\mbox{\boldmath $\theta$}}
\def\balpha{\mbox{\boldmath $\alpha$}}
    \def\bbeta{\mbox{\boldmath $\beta$}}
    \def\bC{\mbox{\boldmath $C$}}		
    \def\bD{\mbox{\boldmath $D$}}
    \def\bc{\mbox{\boldmath $c$}}
    \def\bd{\mbox{\boldmath $d$}}
    \def\bY{\mbox{\boldmath $Y$}}
    \def\by{\mbox{\boldmath $y$}}

\title[Modelling workplace contact networks]{Modelling workplace contact networks: the effects of organizational structure, architecture, and reporting errors on epidemic predictions\footnote{We are grateful to Elena U. Burri and Robert Scherzinger, who helped with the data collection and Roland W. Scholz, who contributed to the diary design.  We are grateful to Ira M. Longini Jr., M. Elizabeth Halloran, Mark S. Handcock, Steven Goodreau, and Martina Morris for providing comments on this work.  The data was made available by ETH Zurich and its collection was funded partly by the Swiss National Science Foundation (Grant 32003B\_127548), partly by ETH basic funding.
This research was supported by a fellowship from the German Academic Exchange Service DAAD to Timo Smieszek (Grant D/10/52328) and by the NIH/NIGMS MIDAS Grant U01-GM070749.  Epidemic simulations were performed on the Biostar computational cluster, made available by the Huck Institutes of the Life Sciences at Pennsylvania State University.  The systems, software, and consulting services for Biostar were provided by ITS Research Computing and Cyber Infrastructure.  Timo Smieszek thanks the UK National Institute for Health Research Health Protection Research Unit (NIHR HPRU) in Modelling Methodology at Imperial College London in partnership with Public Health England (PHE) for funding.  The views expressed are those of the authors and not necessarily those of the NHS, the NIHR, the Department of Health, or Public Health England.}}

 \author[Potter, Smieszek, and Sailer]
        {Gail E. Potter\\ 
California Polytechnic State University, San Luis Obispo, CA, USA\\
Center for Statistics and Quantitative Infectious Disease,\\
Fred Hutchinson Cancer Research Center, Seattle, WA, USA\\  
1 (805) 756-2946     \\
gepotter@calpoly.edu\\[\baselineskip]
Timo Smieszek\\
Center for Infectious Disease Dynamics, Pennsylvania State University\\ \medskip
Modelling and Economics Unit \\
Centre for Infectious Disease Surveillance and Control\\
Public Health England, London, UK\\ \medskip
MRC Centre for Outbreak Analysis and Modelling\\
Department of Infectious Disease Epidemiology\\
Imperial College School of Public Health, London, UK\\ \medskip
NIHR Health Protection Research Unit in Modelling Methodology\\
Department of Infectious Disease Epidemiology\\
Imperial College School of Public Health, London, UK\\
   timo.smieszek@phe.gov.uk\\
[\baselineskip]
Kerstin Sailer\\
The Bartlett School of Graduate Studies, University College London\\
k.sailer@ucl.ac.uk}
\jdate{June 2012}
\pubyear{2012}
\pagerange{\pageref{firstpage}--\pageref{lastpage}}
\doi{S0956796801004857}

\begin{document}

\maketitle

\begin{abstract}
Face-to-face social contacts are potentially important transmission routes for acute respiratory infections, and understanding the contact network can improve our ability to predict, contain, and control epidemics.   Although workplaces are important settings for infectious disease transmission, few studies have collected workplace contact data and estimated workplace contact networks.   We use contact diaries, architectural distance measures, and institutional structures to estimate social contact networks within a Swiss research institute.  Some contact reports were inconsistent, indicating reporting errors.  We adjust for this with a latent variable model, jointly estimating the true (unobserved) network of contacts and duration-specific reporting probabilities.  We find that contact probability decreases with distance, and that research group membership, role, and shared projects are strongly predictive of contact patterns.  Estimated reporting probabilities were low only for 0--5 minute contacts.   Adjusting for reporting error changed the estimate of the duration distribution, but did not change the estimates of covariate effects and had little effect on epidemic predictions.  Our epidemic simulation study indicates that inclusion of network structure based on architectural and organizational structure data can improve the accuracy of epidemic forecasting models.
\end{abstract}


\section{Introduction}

Influenza has a strong impact on public health, with seasonal transmission causing 3-5 million cases of severe illness and up to half a million deaths worldwide each year~\citep{who2009}.  Moreover, recent research has emphasized the ongoing threat of an A(H5N1) ``avian'' influenza pandemic with serious global health consequences~\citep{herfst, russell}.  Large-scale epidemic simulation models have been developed to predict the spread of newly evolved virus strains, as well as compare different intervention strategies: for example, the three models compared in~\cite{halloran}.  These models represent face-to-face human contacts through which influenza can be transmitted, but make assumptions about contact patterns rather than estimating contact network structures from data.  For example, they assume random mixing within mixing groups known to be key to influenza transmission: households, classrooms and schools, workgroups and workplaces.  Several studies have estimated household and school contact networks with the aim of improving model specification for epidemic forecasting models, since these areas are known to be important for disease spread.~\cite{potter} and~\cite{potterhens} estimated household network structures by analyzing data from contact diaries administered in a large-scale multicountry survey of contact patterns~\citep{mossong}.  ~\cite{stehle} describe a face-to-face contact network in a French primary school using proximity sensor data and comment on implications for epidemic interventions.  For example, the extent of class homophily in their network suggests that class-based interventions could control disease spread more efficiently than school closures.  The ongoing Social Mixing And Respiratory Transmission in Schools (SMART) study has recently collected data from students in several Pittsburgh K-12 schools in order to better understand influenza transmission routes within schools~\citep{smart}.  Contact data were collected from proximity sensors, contact diaries, and video-recordings of classrooms and will be used to inform epidemic models and intervention strategies employed by the Centers for Disease Control (CDC).  \cite{salathe} analyze wireless sensor data to describe the contact network in an American high school and demonstrate through simulation studies that using network data to inform interventions can reduce the disease burden.~\cite{potterschool} estimated network structures in a high school using data from contact diaries as well as friendship network data.  Additional studies have used wireless sensors to measure contact patterns and describe network patterns in hospitals (e.g. \cite{vanhems, isella, hornbeck}, as well as at professional conferences (\cite{stehle2, cattuto}), in order to better understand the impact of network structure on disease spread in these settings.  

Workplaces are a potential key area of transmission which has received less attention than households and schools.  Workplace-based interventions have the potential to reduce influenza attack rates, as found by~\cite{kelso}.  Furthermore, interventions exploiting the social structure within workplaces (i.e. regular workgroups), might be substantially more cost-effective, as suggested by modelling studies, than those aimed to the entire organization~\citep{ferguson}.  A better understanding of workplace network structure will help us develop more effective and efficient interventions.

Simulation models used for epidemic prediction rely on detailed demographic and transportation data and vary somewhat in their conceptualization and construction.  Some assume random mixing within mixing groups (homes, schools, classrooms, workplaces, etc.), such as~\cite{chao, ferguson} and~\cite{milne}.  Others create activity schedules based on activity surveys, map these schedules to locations, and assume random mixing within the location (a building or a room within the building), such as \cite{stroud2007, smieszekswiss}, and~\cite{fiozzi}.  While different data sources are used, none of these models use contact data to estimate or validate the workplace contact network structure.  Model specification could be improved by collecting workplace contact data and using it to estimate workplace contact networks.  The network model would ideally include structures relevant to the disease transmission process, but for the sake of parsimony, omit those which have no impact on epidemic dynamics.  We contribute to this area by developing a network model for contacts in workplaces relevant to infectious disease transmission.

We use architectural distances measured between workstations, as well as demographic and social/organizational variables, to model contacts between members of a research institute.  Several papers have explored the relationship between workplace contact and distance between desks, in the context of studying communication patterns.  In their seminal study of seven R\&D labs,~\cite{allen} identified communication patterns as a function of distance.  The closer engineers were located, the higher the probability for communication was. Beyond a distance of 25 to 30 meters between workstations of a pair of engineers, their probability of communicating at least once a week decreased rapidly. These findings were confirmed in more recent studies, where (with one exception) daily face-to-face interaction in eight different knowledge-based organizations seemed to take place at a distance of 15 to 22 metres, depending on the size of the organization, its spatial configuration and office typology~\citep{sailerpenn, sailer}.  Again, longer distances resulted in lower contact frequencies (weekly or monthly) on average.  Most recently, it was argued that detailed architectural distance measures between desks of co-workers provide an important rationale for tie formation in intra-organizational interaction networks. Two actors are more likely to interact with each other when they are closely co-located, even if controlling for structural effects within networks (like transitivity and reciprocity) and organizational effects (perceived usefulness of alter and team affiliation)~\citep{sailermcculloh}.  Our study also uses architectural distance measures to predict contact, but has a somewhat different aim than these.  We focus on durations of face-to-face interactions in order to predict epidemic spread; while the others focused on communication patterns pertaining to workplace productivity.  

This paper contributes two statistical innovations to the area of social network analysis.  The first is in developing models for social networks with valued edges.  A social network may be depicted by representing social actors as \emph{nodes} and the social connections between them by \emph{edges} or \emph{ties}.  We can represent a network mathematically by the \emph{sociomatrix} $\mathbf{Y}$, defined by $Y_{ij}=1$ if there is a tie from person $i$ to person $j$ and $Y_{ij}=0$ otherwise.  We refer to such a network (with $0/1$ edges), as \emph{binary}.  If each edge has a value (for example, the duration of contact between $i$ and $j$), we refer to the network as \emph{valued}.  A commonly used class of network models are exponential family random graph models (ERGMs), a flexible class of model originally developed for binary networks~\citep{strauss}.  These models allows the researcher to include effects such as homophily (the tendency for actors to associate with similar actors), and transitivity (the increased likelihood for a tie between two actors who both have a tie to a common third person).  Previous research has incorporated ERGMs into latent variable models for disease transmission, in which spatial and individual data were observed but the actual contact network is unobserved (e.g.~\cite{groendyke2012}).
Methods have been developed for parameter estimation and simulation of binary networks from ERGMs~\citep{hunter2006, statnet}. ~\cite{krivitsky2012exponential} extends exponential family random graph models (ERGMs) to the case of valued networks, discusses the challenges of model specification and parameter estimation that arise, and applies these techniques to two real social networks.~\cite{hoff2005bilinear} creates a model applicable to valued network data by including a bilinear effect and fixed and random effects in a generalized linear model and proposes a procedure for Bayesian inference.  We explored fitting special cases of the class of models proposed by~\cite{krivitsky2012exponential} to our data and concluded that the most appropriate model for our data required categorizing durations, so creating an ordinal network.  We then use a proportional odds model to model the network of duration categories.  This model for ordinal network data may be applied to networks in a variety of settings.  %We should also note that some work estimating valued networks has also been done with latent space models for networks, in which edges are predicted based on the unobserved position of social actors in a latent social space, as well as covariates. ~\cite{krivitsky2009representing}% latent cluster random effects model and apply it to a network of count data.  In this paper, we model the network of contact durations, a valued network.  

The second statistical advancement in our work is the incorporation of reporting error into our network model.  Recent simulation studies have shown that error in reports of network edges can have a substantial impact on estimation of network parameters~\citep{vznidarvsivc2012non, wang2012measurement, almquist2012random}.  Network researchers frequently note inconsistencies in edge reports, but it is fairly common to discard the information in the inconsistency patterns either by assuming that an edge exists if at least one of the two people involved reported it (e.g., \cite{potterschool}) or restricting analysis to mutually reported edges (e.g., \cite{goodreau}).  When such an approach is taken for inherently symmetric networks, it may result in an underestimate of uncertainty.  For inherently asymmetric networks, it simply results in a loss of information which could be analyzed explicitly.  Some studies have estimated rates of inconsistent reports (e.g., \cite{smieszek2, adamsmoody, hell}) but do not incorporate these estimates in a statistical framework in which they also estimate network effects.  Such a framework is proposed by~\cite{butts2003}, who proposes a hierarchical Bayesian model to jointly estimate posterior distributions of network parameters and probabilities of false negative and false positive tie reports for binary networks.  We instead use a likelihood-based approach, so do not impose additional assumptions implemented by the prior distributions in the Bayesian model.  We also extend our model to ordinal rather than binary networks.  We analyze the same data that~\cite{smieszek2} used to estimate reporting probabilities, but we extend their model to jointly estimate reporting probabilities and network effects.  We validate their findings and explore the impact of adjusting for reporting errors on network estimates and epidemic predictions.  

%~\cite{smieszek2} estimates reporting errors by analyzing inconsistent reports in the contact data analyzed here, finding high reporting probabilities for contacts with long durations, but low reporting probabilities for shorter contacts.  

%\cite{adamsmoody} compute frequencies of concordant ties reported by multiple (two or more) respondents in networks defined by sexual contact, drug-sharing, needle-sharing, and close personal contacts.  Proportions of concordant reports by two people range from 70-94\%.  Their data were collected with a link-tracing design, and reporting inconsistencies between respondents and their nominated alters may be smaller than those between randomly selected respondents.  

%\cite{read} administered longitudinal contact surveys to 27 male and 22 female participants, all of whom were students and staff at University of Warwick.

This paper is organized as follows.  In Section~\ref{subsection:contact} we describe the contact data collected from members of a Swiss research institute, and in~\ref{subsection:distdata} we describe the construction of architectural distance measures computed between desks of these members.  In~\ref{subsection:ergm} we describe the class of network models known as exponential family random graph models (ERGMs), which we employ and expand upon here.  In \ref{subsection:binary} we describe our latent variable model for the binary network of contacts, which jointly estimates ERGM network effects and duration-specific reporting probabilities.  In~\ref{subsection:ordinal} we expand this model to estimate network effects for the network of categorized durations using a proportional odds model, while jointly estimating duration-specific reporting probabilities.  In~\ref{subsection:sims} we describe our epidemic simulation study, which explores whether our model captures the network structures important for disease transmission and compares its predictions to predictions based on random mixing.  We report our estimates for the binary network model in~\ref{subsection:binaryresults} and those for the ordinal network model in~\ref{subsection:ordinalresults}.  We report results for our epidemic simulation study in~\ref{subsection:simresults}.  In Section~\ref{section:discussion} we discuss our findings and make recommendations for future work.

%Finally, we explore through simulation studies whether our model captures the network structures important for disease transmission.  We do this by simulating disease transmission over the observed contact network (where inconsistencies in contact reports have been removed) and comparing disease spread to a simulation over a contact network simulated from our model.  We also perform simulations over simpler versions of our model to assess whether a simpler version would adequately predict disease spread.

\section{Data}
\subsection{Contact survey}
\label{subsection:contact}

Longitudinal contact data and demographic information of the employees of three research groups at a Swiss university were collected using a questionnaire and contact diary approach~\citep{smieszek2}.  The three research groups belonged to one institute, and 66 individuals worked for one or more of the three groups. At least four individuals were absent from work during the entire period of data collection. Fifty individuals completed and returned both questionnaire and diary resulting in a participation rate of at least 80.6\%.   Contact data were collected during five consecutive workdays between Monday, May 17, 2010, and Friday, May 21, 2010.

Participants reported gender, age, research group membership(s), function within the research groups (professor, senior scientist, PhD student, administrative staff), the days on which they were usually in the office, and with whom they shared their office.  Participants also reported all potentially contagious contacts they had with other participants of this study. A potentially contagious contact was defined as either conversation held at $<2$ m distance and with more than ten words spoken, or any sort of physical contact with another individual.  For each contact, participants reported the name of the counterpart, the total contact duration during the entire day (in minutes), and whether the contact was conversational, physical, or both. Contacts were reported separately for each of the five study days. All participants were asked to complete their diaries independently and not to communicate with the other participants about the contents.  For each day analyzed, we omitted from our analysis participants for whom no contacts were reported on that date, assuming this to be an indication of their absence from work.  

\subsection{Architectural distance measures}
\label{subsection:distdata}
\begin{figure}[htbp]
\begin{centering}
  \includegraphics[width=5in,height=5in]{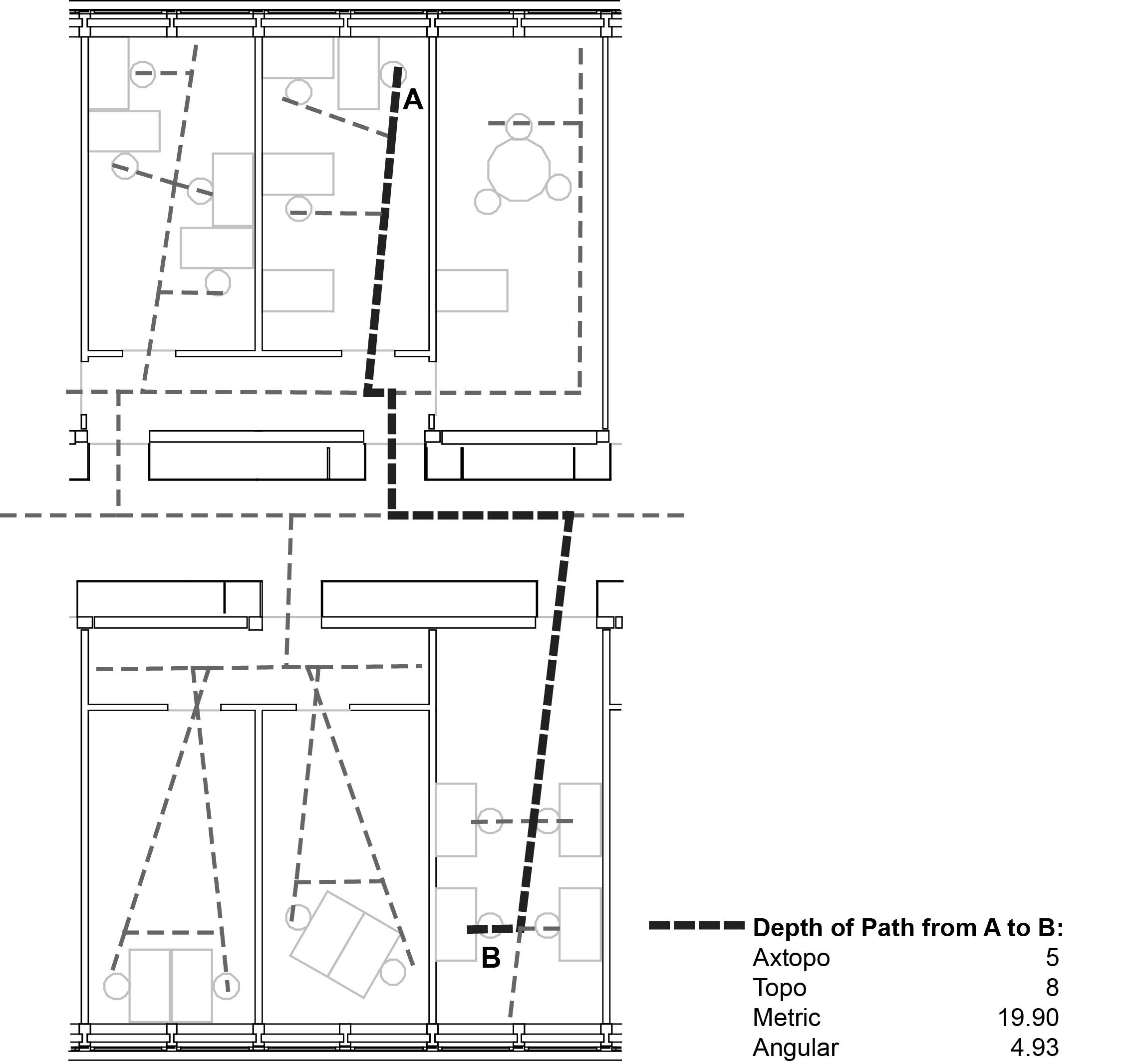}
\end{centering}
\caption{\label{fig:map}  An example office layout with axial topology and four architectural distance measures computed from A to B.}
\end{figure}

This paper uses different ways of measuring the architectural distance between desks of co-workers, as initially introduced by~\cite{sailermcculloh}.  In order to represent distances, a map of lines following possible routes through the office building is drawn using Space Syntax methodologies~\citep{hillierhanson, hillier}. This line map consists of all longest straight lines covering all relevant parts of the office, reaching all individual workstations and minimizing the number of lines and elements needed to go from one space to another (see Figure~\ref{fig:map}). The different floors of the office are linked through the staircases, again with lines representing the potential movement flow of people up and down the flights of the stair. 

With this representation of space as a network of lines, shortest paths can be constructed from one desk to another desk. Based on~\cite{hillieriida} and calculated using the software SEGMEN~\citep{iida}, four different distance measures can be derived from this map for the distance between any desk A and B:
\begin{enumerate}

\item The ``axtopo distance'' between two desks is the number of axes (i.e. full lines) passed on the way from one desk to another.  By convention the root line is not counted.
\item The ``topo distance'' between two desks is the number of segments passed on the way from one desk to another. This is based on each full line broken down into separate segments at each intersection of two lines. Again, the root segment is not counted.
\item The ``metric distance'' between two desks is the total length of the route from one desk to another in walking meters.  The distance is calculated from the center of each segment by convention.
\item The ``angular distance'' between two desks is the sum of changes of direction occurring on the route from one desk to another. The model assigns a 90 degree angle a weight of 1. Thus a route with three 90-degree turns would have angular distance 3. 
\end{enumerate}

Examples of these four distance metrics computed on an office layout are shown in Figure~\ref{fig:map}.
Previous research has shown that these distance measures capture different experiences and notions of distance (perceived distance versus actual distance) and model an existing environment more accurately than the most commonly used Euclidean distances. For more details on this discussion see~\cite{sailermcculloh}.

\section{Methodology}
\subsection{Exponential family random graph models}
\label{subsection:ergm}
We use exponential random graph models, following the example of~\cite{sailermcculloh} to estimate the effects of individual attributes and architectural distances on contact patterns.  We also expand on ideas set forth in~\cite{smieszek2}, who use these data to estimate the probability of reporting an existing contact.  We develop a model to estimate ERGM parameters for the true network of contacts, statistically correcting for measurement error and inconsistencies in reporting.  We do this by expressing the likelihood of the network of reported contacts in terms of the network of true contacts and the probability of reporting a contact.  Parameter estimates are then obtained through maximum likelihood estimation. 

We first define some network terminology and notation.  We can depict the network by representing the social actors by nodes and the contacts between them with edges or ties.  A pair of actors is called a dyad.  We represent the observed contact network by a sociomatrix, $\bC$, a square matrix with as many rows as there are people in the network, where $C_{ij}=1$ if $i$ reports contact to $j$ and $0$ otherwise.  We represent the sociomatrix of true contacts by $\bY$, so $Y_{ij}=1$ if $i$ and $j$ actually made contact, regardless of whether that contact was reported, and $Y_{ij}=0$ if no contact was made.  Because of inconsistencies in reporting, $\bC$ is an asymmetric matrix.  However, the sociomatrix of true contacts, $\bY$, is symmetric because contact by our definition is symmetric.  Let the four duration categories 0--5, 6--15, 16--60, or 60--480 minutes be denoted by $d_k,$ for $k\in \{1,2,3,4\} $, and let $p_k$ denote the probability of reporting an existing contact of duration $d_k$.  We assume the reporting probability is independent of the contact probability for any two actors.  We assume, as in~\cite{smieszek2}, that the events that different respondents report existing contacts are independent, and that no contacts were fabricated.  We also assume that the reporting probability does not depend on covariates which may predict contact patterns.  This assumption is comparable to a \emph{missing completely at random} (MCAR) assumption if reporting errors are viewed as missing data.  We explored modeling such a dependency (discussed below) but found this not to improve our model.  

We would like to jointly estimate $\mathbf{p}$ and $\bY$.  Since $\bY$ is unobserved, we will use a latent variable model for estimation.  We will express the likelihood of our observed data $\bC$ in terms of $\bY$ and $p$ and compute the maximum likelihood estimate.

We use exponential family random graph models (ERGMs) to model the true contact network $\bY$.  An ERGM takes the form 
\[
P(\bY=\by|\btheta) = \frac{
e^{ \btheta^T \mathbf{g(y)} }}{\kappa (\mathcal{\btheta, Y})}
\]
Above, $\mathcal{Y}$ denotes the set of all networks of this size, ${\kappa (\mathcal{\btheta, Y})}$ is a normalizing constant ensuring that the probability distribution sums to one, $\btheta$ is a vector of parameters, and $\mathbf{g(y)}$ is a vector of network statistics capturing network structures we want to estimate.  For example, $\mathbf{g(y)}$ may include an edges term for a density effect, the number of contacts between members of the same sex for a mixing effect, or a triangles term to capture transitivity (the increased likelihood of two people who have mutual friends to be friends).  The parameters $\btheta$ are estimated with the maximum likelihood estimate (MLE).  In general the normalizing constant does not have an analytic form, so the MLE is approximated with an MCMC procedure~\citep{snijders2002markov}.  

The choice of statistics in $\mathbf{g(y)}$ specifies the model.  Some ERGMs are \emph{dyad-independent}, which means that the event of contact occuring on one dyad is independent of contact patterns on other dyads.  In dyad-independent models, contact behavior is characterized by individual-level and dyadic attributes, and the MLE may be estimated with logistic regression rather than MCMC.  In \emph{dyad-dependent} models, $\mathbf{g(y)}$ includes dependency terms, such as the number of triangles.

We included the following statistics in our ERGM:
\begin{enumerate}
\item The number of edges (a density effect)
\item Two terms to estimate sociality effects for each research group: the number of contacts made by members of research group 1 and the count for group 2.  Group 3 is used as the reference group, so these terms estimate how much more social members of groups 1 and 2 are than members of group 3.
\item The number of contacts made between members of the same research group, estimating a preference to contact others in the same group.  This effect is distinct from the previous one, because while some groups may be more social than others, their contacts may occur in different ratios of between vs. within-group contacts.  
\item The total distance between members making contact.  We fit four separate ERGMS with four separate distance metrics.  
\item Sociality effects by gender: the number of contacts made by females.
\item Gender homophily: the number of same-gender contacts.
\item Class homophily: the number of contacts between members of the same function (graduate students, postdocs, or administrative staff).  No professors participated in the survey.
\item The total number of contacts between people on the same floor.  People may be more inclined to contact others on the same floor.
\item Shared-projects homophily: the total number of contacts between people who work on the same projects together (weighted as 1 or 2, depending on whether they are mutually reported). 
%\item The sum of integration measures of people in contact.  People with higher integration measures %may be more likely to make contact.  
\end{enumerate}

The ERGMs we selected are dyad-independent, in which case the likelihood of the actual network is equivalent to logistic regression with dyads as the dependent variable.  The assumption of dyad-independence is a strong one, since additional clustering may be present in the network which is not explained by the various mixing effects included in our model.  Adjusting for reporting errors with a dyad-dependent ERGM would be extremely complex mathematically and computationally, so we chose to begin with the simpler dyad-independent models.  We express the probability distribution of one dyad as:

\[
\textnormal{logit}(P(Y_{ij}=1)) = \beta_0 + \beta_1 (1_\textnormal{[i in group 1]} +  1_\textnormal{[j in group 1]} )+
\]
\[
\beta_2 (1_\textnormal{[i in group 2]} +  1_\textnormal{[j in group 2]} )+
\beta_3 1_{[\textnormal{i, j in same group}]} + 
\]
\[
\beta_4 \textnormal{distance}(i,j) + 
\beta_5 (1_{[\textnormal{i female}]}+1_{[\textnormal{j female}]})+
\beta_6 1_{\textnormal{[i,j same sex]}} \]
\[
\beta_7 1_{\textnormal{[i,j same function]}} +
\beta_8 1_{\textnormal{[i,j on same floor]}} +
\]
\[
 \beta_9 ( 1_{\textnormal{[i reports shared projects with j]}} +
 1_{\textnormal{[j reports shared projects with i]}} )
\] 
When contacts are symmetric and dyads are independent, we obtain:
\[
P(\bY=\by) = \prod_{i=1}^n \prod_{j=i+1}^n (P(Y_{ij}=y_{ij}))
\]

\subsection{Likelihood of reported contacts, duration-specific reporting probabilities}
\label{subsection:binary}
Next we express the likelihood of the reported contacts, $\bC$.  We expect reporting probability to vary with duration of contacts;~\cite{smieszek2} found that shorter contacts were more likely to be forgotten.  In this section we jointly model the reported contacts and durations, and we estimate separate reporting probabilities for each duration category: 0--5, 6--15, 16--60, or 60--480 minutes. We denote the four duration categories by $d_k,$ for $k\in \{1,2,3,4\} $.  As in~\cite{smieszek2}, when two participants reported different durations for the same contact, we assume that the longer duration is the correct report.  We also assume that duration of contact does not depend on individual or dyadic attributes, and we again assume independence in contact (and durations) between dyads, conditional on the effects in our model.  Let $\gamma_k$ denote the probability of an existing contact having duration $d_k$, so $\gamma_k = P(D_{ij}=d_k | Y_{ij}=1)$.  Let $\bD$ denote the matrix of contact durations, after removing inconsistencies in duration reports, so $\bD$ is a symmetric matrix.   By applying our assumptions (including dyad independence), rules of conditional probability, and the Law of Total Probability, we find that the joint likelihood of $D_{ij}$ and $C_{ij}$ is:
\[
P(C_{ij}=1, C_{ji}=1, D_{ij}=d_k)=\gamma_k P(Y_{ij}=1)p_k^2
\]
\[
P(C_{ij}=0, C_{ji}=1, D_{ij}=d_k)=\gamma_k P(Y_{ij}=1)p_k(1-p_k)
\]
\[
P(C_{ij}=0, C_{ji}=0)= P(Y_{ij}=0) + \sum_{k=1}^4 \gamma_k P(Y_{ij}=1) (1-p_k)^2 
\]

Then the probability of observing the reported contact network is found by using the above equations to express the probabilities in the following formula:
\[
P(\bC=\bc, \bD=\bd) = \prod_{i=1}^n \prod_{j=i+1}^n P(C_{ij}=c_{ij}, C_{ji}=c_{ji}, D_{ij}=d_k)
\]

We maximized the log likelihood using R software with the {\tt trust} function in R~\citep{trust, R2011}.  This optimization method requires gradient and Hessian functions of the log likelihood as input values, and we approximated these with the {\tt grad} and {\tt hessian} functions in the {\tt numDeriv} package in R~\citep{numDeriv}.  The optimization routine returns the parameter vector maximizing the log likelihood as well as the value of the Hessian at the MLE.  We computed standard errors by inverting the negative of the Hessian (the observed Fisher information matrix).  

\subsection{Likelihood of contact durations with duration-specific reporting probabilities}
\label{subsection:ordinal}

The duration of each contact may depend on individual and dyadic attributes.  For example, short contacts may occur more frequently between those who share an office.  The idea is that if one travels an extra distance to contact another person, they are likely to make a longer contact.  In the previous section we assumed that durations were uniformly distributed; in this section we refine that model to estimate duration from known attributes.

\begin{figure}[htbp]
\begin{centering}
  \includegraphics[width=4in,height=4in,keepaspectratio]{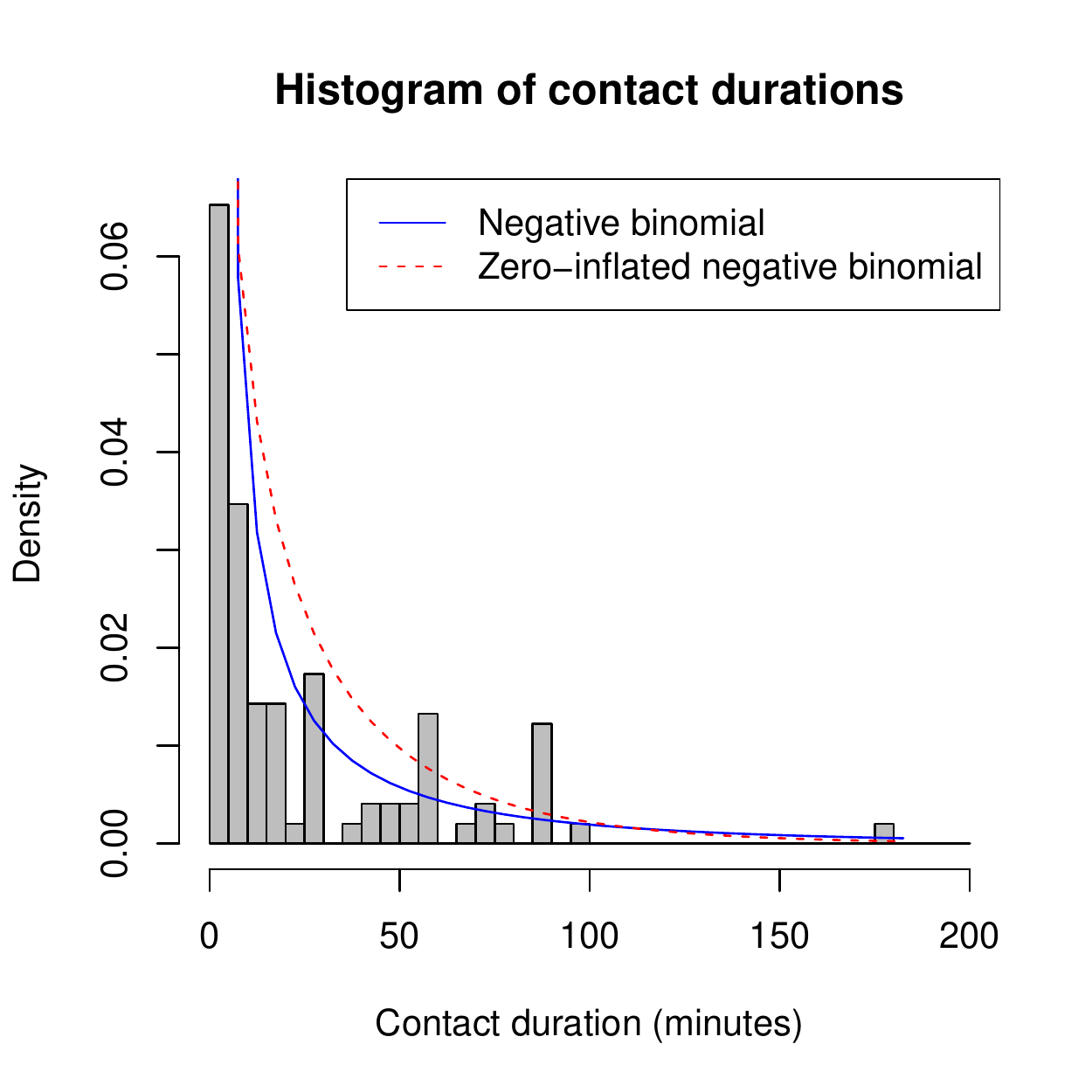}
\end{centering}
\caption{\label{fig:durdist}  The observed distribution of contact durations, with negative binomial and zero-inflated negative binomial fitted models.  Contacts
with zero duration (i.e., non-contacts) comprised 92\% of all dyads and are omitted from the graph.}
\end{figure}

In order to do this, we need to create a model for the probability distribution of the duration matrix and derive the expression for its likelihood.  Then we will express the joint likelihood of the true duration matrix and the reported duration matrix as in previous sections, and maximize the log likelihood function with {\tt trust}.  Our reported durations have a large number of zeroes and are overdispersed. The mean of the nonzero duration reports is 26 minutes, and variance 987. We could use a generalized linear model to estimate the mean of the duration distribution as a function of covariate values~\citep{glm}. For this approach, we considered a negative binomial distribution and a zero-inflated negative binomial distribution (fits shown in Figure~\ref{fig:durdist}).  Our actual distribution of contact durations has spikes at 30, 60, and 90 minutes, either because people tended to round their durations to these values or because based on common meeting lengths, these values are actually more frequent.  The parametric forms we considered did not capture this phenomenon, so we instead categorized duration in order to avoid imposing assumptions on the duration distribution.  

We used two methods to estimate probabilities of a duration falling into each category as a function of covariates.  The first approach was multinomial logistic regression, which has no distributional assumptions but does not take advantage of the fact that duration categories are ordered.  The second method was a proportional odds model, which does exploit the ordering, but imposes an additional assumption.  We found the proportional odds model to be more appropriate for this data set.  The multinomial model created some estimation problems due to the large number of parameters and some cases of very small cell counts. While the multinomial model allows additional flexibility, the results did not show strong evidence that effects varied by duration category. Since the additional flexibility did not yield extra insight, we decided that the proportional odds model is preferable, so we restrict our attention here to that model. A detailed description of the multinomial model and its performance is included in the supplementary material.

The proportional odds model is defined by:
\[
\textnormal{logit(P($D_{ij}\le d_k$))} = \alpha_k - \mathbf{\bbeta}^T\bf{x}
\]
While the multinomial model estimates a separate vector of covariate effects for each duration category, here a single vector of covariate effects is estimated.  The indexing variable $k$ takes values from zero to four, for each possible duration category (0, 1-5, 6-15, 16-60, $>$60).  

This model satisfies:
\[
\textnormal{logit(P(}D_{ij} > d_k | \mathbf{x_1})) - \textnormal{logit(P(}D_{ij} > d_k | \mathbf{x_2})) = \mathbf{\bbeta}^T(\bf{x_1}-\bf{x_2}),
\]
i.e., the the log cumulative odds ratio is a fixed linear combination of the differences in the covariate values, and this linear combination is the same for each category of the outcome variable.  The probability of observing duration category $k$ for contact between $i$ and $j$ is:

\[
P(D_{ij}=d_1 | X_{ij}= \mathbf{x}) = 
\frac{  e^{\alpha_1-\bbeta^T \mathbf{x}}  }{ 1+ e^{\alpha_1-\bbeta^T \mathbf{x}}   }
\]

\[
P(D_{ij}=d_k | X_{ij}=\mathbf{x}) = \frac{  e^{\alpha_k-\bbeta^T \mathbf{x}}  }{ 1+e^{\alpha_k-\bbeta^T \mathbf{x}}   }-
\frac{  e^{\alpha_{k-1}-\bbeta^T \mathbf{x}}  }{ 1+ e^{\alpha_{k-1}-\bbeta^T \mathbf{x} }   } , \textnormal{ for } k=1, 2, 3, 4
\]

\[
P(D_{ij}=d_5 | X_{ij}=\mathbf{x}) = 1-\sum_{k=1}^4 P(D_{ij}=d_k | X_{ij}=\mathbf{x}) 
\]

By applying our assumptions, rules of conditional probability, and the Law of Total Probability, we find that the joint likelihood of $\bD$ and $\bC$ is:
\[
P(C_{ij}=1, C_{ji}=1, D_{ij}=d_k)= P(D_{ij}=d_k)p_k^2
\]
\[
P(C_{ij}=1, C_{ji}=0, D_{ij}=d_k)=P(D_{ij}=d_k) p_k(1-p_k)
\]
\[
P(C_{ij}=0, C_{ji}=0, D_{ij}=0)= P(D_{ij}=0) + \sum_{k=2}^5 P(D_{ij}=d_k) (1-p_k)^2 
\]

Then the probability of the observed data is:
\[
P(\bC=\bc, \bD=\bd) = \prod_{i=1}^n \prod_{j=i+1}^n P(C_{ij}=c_{ij}, C_{ji}=c_{ji}, D_{ij}=d_k)
\]
We maximize the log likelihood to estimate $\balpha$, $\bbeta$, and $\bf{p}$ using the {\tt trust} function in R and computed standard errors by inverting the Fisher information matrix~\citep{trust}.

We tested the proportionality assumption by dichotomizing the outcome variable at each of the duration cutpoints, fitting separate logistic regression models, and comparing odds ratio estimates from the different models.  In testing the assumption, we did not jointly estimate reporting probabilities together with covariate effects because extending our model is not straightforward when contact duration is dichotomized.  Instead we assumed that contact occurred if either or both members reported it.  These results are included in the appendix and indicate that the proportional odds assumption is reasonable.

We considered modelling the reporting probability as a function of covariates.  For example, older people may be more likely to forget contacts than younger people.  Exploratory data analysis revealed that members are less likely to forget contacts to those who work on a different floor, but did not show evidence for other covariate effects on reporting probability.  Among those on the same floor, 17\% of reports were consistent, while 82\% of contact reports between members on different floors were consistent.  We extended the proportional odds model to include a floor effect on reporting probability.  We estimated the log odds ratio of reporting probability for different to same floors, and assumed the same odds ratio for different contact durations.  However the estimated log odds ratio was not statistically significant (95\% C.I. [-1.23, 0.49]), so we decided to omit this effect from our final model.  Our power is limited because there were only eleven contact reports between members on different floors. 

\subsection{ Simulation model of epidemics}
\label{subsection:sims}
Our epidemic simulations were based on an individual-based model of influenza spread used in previous publications~\citep{salathe, smieszek2013low}.  The model is a stochastic SEIR model with simulation time steps of half a day. We assume that infection is introduced by one randomly chosen index case at the beginning of a simulation run and that, after the initial introduction, there are no further introductions from outside. 

When a susceptible individual has contact with one or more infectious individuals, the probability to switch from the susceptible to the exposed state is  $1 - (1 - \psi)^w$, where $\psi$ is the probability of transmission per minute and $w$ is the accumulated contact time the susceptible individual has spent with infectious individuals during the entire half-day at work. Previous work estimated the transmission probability for an influenza outbreak that occurred on a plane to be $\psi = 0.009$ min$^{-1}$~\citep{moser}.  We used this as an initial input for our simulations.  As influenza strains vary in their infectiousness, we performed additional simulations using fifteen different infectivity parameters, ranging from $\psi = 0.009$ min$^{-1}$ to $\psi=0.135$ min$^{-1}$. The duration of the exposed state follows a Weibull distribution with an offset of half a day; the power parameter is 2.21, the scale parameter is 1.10~\citep{ferguson2005strategies}.  After the exposed state, each individual remains in the infectious state for exactly one time step, and then withdraws to the home, so is removed from the workplace population. 

A week in our simulation model consists of 14 timesteps: five workdays with each one time step at the workplace and one time step at home as well as two weekend days with in total four time steps at home. Since workplace transmission can only occur during the five time steps at work, any exposed individual turning infectious at home, will not be able to pass on the infection to colleagues at work. Unlike previous applications of this model ~\citep{salathe, smieszek2013low}, we assume here that the initial case will become infectious during one of the five workday time steps of a simulation week, which is determined randomly for each run.

In order to assess the adequacy of our model to represent the actual network structure, we compare epidemic simulations based on simulated contact networks from our model (see Table~\ref{tab:comparePOM}) to those based on the original data collected on the Monday of the study week.  For simulations based on the original data, we assume that contact occurred if it was reported by one or both members.  We also compare these two results to simulations based on a version of our model which does not adjust for inconsistent reports, and which uses the ``standard assumption'' which converts discordant ties to concordant ones.  

We compare our network-based epidemic simulation results to a random mixing model, commonly used in epidemic predictions.  To make a fair comparison, the random mixing model has the same total number of contacts (on average) as a network simulated from our model, and durations in the random mixing model are all equal to the mean duration.  We also perform epidemic simulations based on subsets of our model, shown in Table~\ref{tab:subsets}, in order to assess whether a more parsimonious model would sufficiently represent the network structure relevant to epidemic spread.  Finally, we perform simulations based on contact networks generated by random reshuffling of the edges generated from one simulation of the full model. To this end, two pairs of nodes were randomly chosen, and their contact duration (including no contact operationalized as contact of zero duration) was swapped. This was repeated 100000 times.  The reshuffling model preserves the distribution of edge duration alone, so testing the effect of this single network structure.  

We generated 500 different realizations for all contact network models and computed 10000 individual simulation runs for each realization.

\section{Results}

Table~\ref{tab:descriptives} shows descriptive statistics of our data set.  There are equal numbers of men and women, and most members are graduate students or postdocs.  No professors responded to the survey.  One member belonged to all three groups and two members belonged to both groups one and two.  

\begin{table}[ht]
\begin{center}
\caption{Descriptive statistics of 50 members of the Swiss research institute}
\begin{oldtabular}{lllll}
\hline
\multicolumn{1}{l}{Variable\textcolor{white}{space}}& \multicolumn{3}{c}{Mean(SD) or frequency table}\\ 
\hline
Age & 31.7 (6.6)  &&\\
\\
Sex  &25 Male \\
&25 Female & \\
\\
Role & 15 Postdoctoral fellows \\
& 24 Graduate students \\
& 6 Administrative staff\\
\\
Group		& 24 Group 1 \\
& 19 Group 2 \\
& 11 Group 3\\
\hline
\label{tab:descriptives}
\end{oldtabular}
\end{center}
\end{table}

\subsection{Results for models for binary contact network}
\label{subsection:binaryresults}
Table~\ref{tab:duration} shows parameter estimates for our model for the binary contact network fit with four different distance metrics.  The best-fitting model according to the AIC is the model including angular distance.  Estimates for this model are interpreted as follows: the odds of a contact stemming from a member of group 2 is $e^{0.38}=1.46$ times the odds of contact stemming from a member of group 3 (the reference group), controlling for other terms in the model. The odds of contact is increased by a factor of $e^{3.48}=32$ if the two people belong to the same research group, controlling for other effects.  For each additional unit of angular distance between two workstations, the odds of contact decreases by a factor of $e^{-0.27} = 0.76$, controlling for other effects in the model.  The odds of contact is increased by a factor of $e^{0.14}=1.15$ if it stems from a female rather than from a male, although this effect is not significant.  The odds of contact increases by a factor of $e^{1.44}=4.22$ if they share work projects.  Effects in the other models are interpreted similarly.  The coefficient for shared projects in the metric distance and axtopo distance models is effectively infinite, meaning that once controlling for other effects in the model, those who share projects always make contact.  The angular and axtopo distance models perform similarly, because the construction of these distance metrics is similar.  The metric and topo models perform similarly for the same reason.

\begingroup

%\newgeometry{left=.5in, right=.5in}
\begin{table}[ht]
\begin{center}
\caption{Coefficient estimates for logistic regression models of binary contact network with four different distance metrics.}
\footnotesize
\label{tab:duration}
\setlength{\tabcolsep}{4pt}
\begin{tabular}{lrlrlrlrl}
 &\multicolumn{8}{c}{Architectural distance metric included in ERGM}\\ 
\hline
Estimate & Metric & & Angular & & Topo & & Axtopo & \\ 
  \hline
%Intercept & -5.59 (1.22) & *** & -2.27 (1.21) & . & -5.93 (1.33) & *** & -2.13 (1.34) &  \\ 
  Group 1 & -0.34 (0.24) &  & 0 (0.23) &  & -0.35 (0.24) &  & -0.04 (0.22) &  \\ 
  Group 2 & 0.15 (0.23) &  & 0.38 (0.23) & . & 0.14 (0.22) &  & 0.40 (0.23) & . \\ 
  Group mixing & 3.75 (0.55) & *** & 3.48 (0.48) & *** & 3.77 (0.54) & *** & 3.44 (0.48) & *** \\ 
  Distance & 0.01 (0.02) &  & -0.27 (0.11) & * & 0.04 (0.05) &  & -0.23 (0.10) & * \\ 
  Female & 0.26 (0.25) &  & 0.14 (0.24) &  & 0.26 (0.25) &  & 0.14 (0.24) &  \\ 
  Role mixing & 0.68 (0.36) & . & 0.60 (0.34) & . & 0.70 (0.36) & . & 0.63 (0.34) & . \\ 
  Gender Mixing & -0.22 (0.31) &  & -0.23 (0.30) &  & -0.21 (0.31) &  & -0.23 (0.30) &  \\ 
  Floor & 1.15 (0.59) & . & -0.69 (0.81) &  & 1.38 (0.70) & . & -0.93 (0.94) &  \\ 
  Shared projects & 19.02 (NA) &  & 1.44 (0.53) & ** & 21.46 (NA) &  & 1.47 (0.54) & ** \\ 
\\
 \multicolumn{2}{l}{Reporting probability}\\
0--5 & 0.48 [0.36, 0.60] &  & 0.56 [0.41, 0.69] &  & 0.48 [0.36, 0.60] &  & 0.55 [0.41, 0.69] &  \\ 
  6--15 & 0.96 [0.84, 0.99] &  & 0.96 [0.84, 0.99] &  & 0.96 [0.84, 0.99] &  & 0.96 [0.84, 0.99] &  \\ 
  16--60 & 0.93 [0.83, 0.98] &  & 0.93 [0.83, 0.98] &  & 0.93 [0.83, 0.98] &  & 0.93 [0.83, 0.98] &  \\ 
  61-480 & 1.00 [0, 1.00] &  & 1.00 [0, 1.00] &  & 1.00 [0, 1.00] &  & 1.00 [0, 1.00] &  \\ \\
 \multicolumn{2}{l}{Duration category}\\
0--5 & 0.49 [0.39, 0.59] &  & 0.45 [0.35, 0.55] &  & 0.49 [0.39, 0.59] &  & 0.46 [0.35, 0.56] &  \\ 
  6--15  & 0.18 [0.12, 0.25] &  & 0.20 [0.12, 0.27] &  & 0.18 [0.12, 0.25] &  & 0.20 [0.12, 0.27] &  \\ 
  16--60  & 0.24 [0.16, 0.31] &  & 0.25 [0.17, 0.33] &  & 0.24 [0.16, 0.31] &  & 0.25 [0.17, 0.33] &  \\ 
  61-480  & 0.09 [0.04, 0.14] &  & 0.09 [0.04, 0.15] &  & 0.09 [0.04, 0.14] &  & 0.09 [0.04, 0.15] &  \\ 
\hline \\
  AIC & 796 &  & 789 &  & 795 &  & 790 &  \\ 
   \hline

\end{tabular}
\end{center}
Significance levels: *** $=p<0.001$; ** $=p<0.01$; * $=p<0.05$; ``.'' $=p<0.10$
\end{table}
%\restoregeometry
\endgroup

The four models estimate similar reporting probabilities and duration categories.  The reporting probability estimates were nearly identical to those obtained by~\cite{smieszek2}, validating our method.  Table~\ref{tab: reporting} in the Appendix compares both sets of estimates.

Table~\ref{tab:durdist} compares the estimated duration distribution from the model with angular distance to the duration distribution of reported contacts.  Our model necessarily estimates higher numbers of contacts than are observed, since some non-contacts are attributed to reporting errors but no contacts are considered erroneously reported.  Since a much lower reporting probability (0.56) is estimated for 0--5 minute contacts than for longer duration contacts (0.93--1.00), our model estimates a higher proportion of 0--5 minute contacts than what is observed.

\begin{table}[ht]
\begin{center}
\setlength{\tabcolsep}{6pt}
\caption{Duration distribution estimates from our model compared to the observed duration distribution of reported contacts.}
\label{tab:durdist}
\begin{tabular}{lrrr}
  \hline
 Duration& Our estimate of & Distribution of \\
 category & true distribution & reported contacts\\% (adjusting for reporting error) & (no error adjustment)\\
\hline
0--5 & 0.45 [0.35, 0.55] &   0.41 [0.25, 0.56]\\ 
6--15& 0.20  [0.12, 0.27]   & 0.22 [0.00, 0.47] \\ 
16--60 & 0.25  [0.17, 0.33]  & 0.28 [0.06, 0.49] \\ 
$>60$ & 0.09 [0.04, 0.15]   & 0.10 [0.00, 0.49] \\ 
\hline
\end{tabular}
\end{center}
\end{table}

\subsection{Results for proportional odds models for network of contact durations}
\label{subsection:ordinalresults}

The proportional odds model with angular distance was also best according to the AIC, so we present only that one here, although the others are included in the Appendix in Table~\ref{tab:POM01}.  Table~\ref{tab:comparePOM} compares estimates from our model to those from a model which does not adjust for reporting errors (the ``standard model'') but instead assumes that contact occurred between two people if and only if at least one of the two reported it.  Once again, duration distributions differ slightly (though not significantly), but effect estimates are nearly identical.  The standard errors for our model are larger than those for the standard model, because they incorporate the additional uncertainty contributed by reporting errors.  Coefficients for this model are interpreted as follows: The odds of duration being greater than a certain category increases by a factor of $e^{3.42}=30.6$ when two members are in the same group.  The odds of duration being greater than a specified category decreases by a factor of $e^{-0.22}=0.80$ with each additional unit of angular distance between their workstations.

\begin{table}[ht]
\begin{center}
\caption{Coefficient estimates for proportional odds model, compared to estimates using standard assumption, angular distance measure}
\label{tab:comparePOM}
\setlength{\tabcolsep}{6pt}
\begin{tabular}{lrlrlrl}
  \hline
 & Our model& & Standard model&\\ 
  \hline
Intercepts &Est. (SE) &&Est. (SE) &\\
  \hline
Duration = 0 & 2.65 (1.04) & * & 2.81 (0.72) & *** \\ 
Duration $\le$ 5 & 3.87 (1.04) & *** & 3.76 (0.72) & *** \\ 
Duration $\le$ 15 & 4.58 (1.04) & *** & 4.47 (0.72) & *** \\ 
Duration $\le$ 60 & 6.22 (1.08) & *** & 6.10 (0.74) & *** \\ 
\hline
Coefficients\\
\hline
  Group 1 & -0.18 (0.20) &  & -0.16 (0.14) &  \\ 
  Group 2 & 0.11 (0.19) &  & 0.12 (0.13) &  \\ 
  Group mixing & 3.42 (0.45) & *** & 3.34 (0.31) & *** \\ 
  Distance & -0.22 (0.08) & ** & -0.22 (0.06) & *** \\ 
  Female & 0.31 (0.21) &  & 0.29 (0.14) & * \\ 
  Role mixing & 0.60 (0.29) & * & 0.58 (0.20) & ** \\ 
  Gender mixing & -0.18 (0.26) &  & -0.17 (0.18) &  \\ 
  Floor & -0.09 (0.68) &  & -0.13 (0.47) &  \\ 
  Shared projects & 1.06 (0.28) & *** & 1.07 (0.19) & *** \\ 
\hline
\end{tabular}
\end{center}
Significance levels: *** $=p<0.001$; ** $=p<0.01$; * $=p<0.05$;  ``.'' $=p<0.10$
\end{table}

Table~\ref{tab:subsets} shows estimates for subsets of the full model.  We performed epidemic simulations over these more parsimonious models to compare them to the full model and assess whether one of them adequately captured all network structure relevant for epidemic predictions.

\begin{table}[ht]
\footnotesize
\begin{center}
\caption{Coefficient estimates for proportional odds model and subsets of model, angular distance measure}
\setlength{\tabcolsep}{4pt}
\label{tab:subsets}
\begin{tabular}{lrlrlrlrlrl}
  \hline
 & Model 1 &  & Model 2 &  & Model 3 &  & Model 4&  &Full Model &  \\ 
  \hline
  0 & -0.48 (0.71) &  & 3.93 (0.39) & *** & 3.92 (0.39) & *** & 2.67 (1.04) & * & 2.65 (1.04) & * \\ 
  1-5 & 0.45 (0.70) &  & 4.95 (0.40) & *** & 4.89 (0.40) & *** & 3.88 (1.05) & *** & 3.87 (1.04) & *** \\ 
  6-15 & 1.03 (0.69) &  & 5.56 (0.42) & *** & 5.45 (0.41) & *** & 4.58 (1.05) & *** & 4.58 (1.04) & *** \\ 
  16-60 & 2.49 (0.72) & ** & 7.10 (0.49) & *** & 6.89 (0.48) & *** & 6.23 (1.08) & *** & 6.22 (1.08) & *** \\ 
\hline
Coefficients\\
\hline
  Group 1 &  &  &  &  &  &  &  &  & -0.18 (0.20) &  \\ 
  Group 2 &  &  &  &  &  &  &  &  & 0.11 (0.19) &  \\ 
  Group mixing &  &  & 3.65 (0.41) & *** & 3.87 (0.41) & *** & 3.37 (0.44) & *** & 3.42 (0.45) & *** \\ 
  Shared projects &  &  & 1.31 (0.27) & *** &  &  & 0.98 (0.27) & *** & 1.06 (0.28) & *** \\ 
  Distance & -0.36 (0.07) & *** &  &  &  &  & -0.22 (0.08) & ** & -0.22 (0.08) & ** \\ 
  Floor & 0.69 (0.47) &  &  &  &  &  & 0.14 (0.62) &  & -0.09 (0.68) &  \\ 
  Female &  &  &  &  &  &  & 0.17 (0.18) &  & 0.31 (0.21) &  \\ 
  Role mixing &  &  &  &  &  &  & 0.61 (0.29) & * & 0.60 (0.29) & * \\ 
  Gender mixing &  &  &  &  &  &  & -0.20 (0.26) &  & -0.18 (0.26) &  \\ 
\hline
\\
AIC & 894 &  & 805 &  & 803 &  & 771 &  & 773 &  \\ 
\hline
\end{tabular}
\end{center}
\end{table}

\subsection{Epidemiological properties of the contact networks and results of the epidemic simulations}
\label{subsection:simresults}

\subsubsection{Correlation of individual epidemiological importance}

We analyzed to what extent the individual epidemiological importance of the members of the workplace population is correlated among the different contact network models. Here, epidemiological importance is operationalized as the mean expected number of cases generated by a specific individual, given that all of its workplace contact partners are fully susceptible. We calculated Kendall's $\tau$ as a robust, non-parametric measure of correlation. If we compare the rank order of the individuals (according to the expected number of cases generated) for two different types of contact networks, then $\tau = 1$ means that the rank order is identical. Contrary, $\tau = 0$ indicates that the ranks of individuals are unrelated between two networks. All other values of Ï„ are as easy to interpret since the odds ratio of concordant to discordant pairs of observations is given by$\frac{1+\tau}{1-\tau}$\citep{tau}.  Table~\ref{tab:X1} shows Kendall's $\tau$ for all pairs between (i) the original data and all other contact networks as well as (ii) the full network model and all other contact networks. 

\begin{table}[ht]
\begin{center}
\caption{Kendall's $\tau$ computed for the rank order of individuals (according to the expected number of cases generated) between various models}
\label{tab:X1}
\setlength{\tabcolsep}{6pt}
\begin{tabular}{lcrrrrrr}
\hline
%&&&&&&&\\
&	$\psi$ (1/min)	& Full &	Standard &	Model 1 &	Model 2	&Model 3	&Model 4\\
&& model & model & &&&\\
\hline
Original	&0.009	&0.33	&0.36	&0.21	&0.14	&0.04	&0.37\\
&	0.045	&0.41	&0.40	&0.28	&0.19	&0.17	&0.40\\
&	0.090	&0.44	&0.43	&0.27	&0.19	&0.26	&0.43\\
&	0.135	&0.48	&0.46&	0.29	&0.19&	0.22	&0.47\\
\hline
&&&&&&&\\
Full model & 0.009	&n/a&	0.95&	0.45&	0.39&	0.42	&0.78\\
&	0.045&	n/a	 &0.95	&0.47	&0.38&	0.43	&0.78\\
&	0.090&	n/a&	0.97&	0.47&	0.36&	0.55&	0.80\\
&	0.135&	n/a&	0.97&	0.50&	0.38&	0.52&	0.82\\
\hline
\end{tabular}
\end{center}
\end{table}

The values are interpreted as follows: when comparing the individual epidemiological importance for the full model and the original data, the odds of concordant rank orders for any pair of observations was approximately twice the odds of a discordant one (since $\frac{1+0.33}{1-0.33}=1.99$).  Comparing model 1 and the original data, the odds of a concordant pair was  approximately 1.5 times that of a discordant pair. A comparison of the full and the standard model resulted in an odds ratio of 39, indicating that the correction for underreporting had very little effect on the individuals' ranking. 

\subsubsection {Simulated epidemics}

\begin{figure}
\begin{centering}
  \includegraphics[width=5.9in,height=4in]{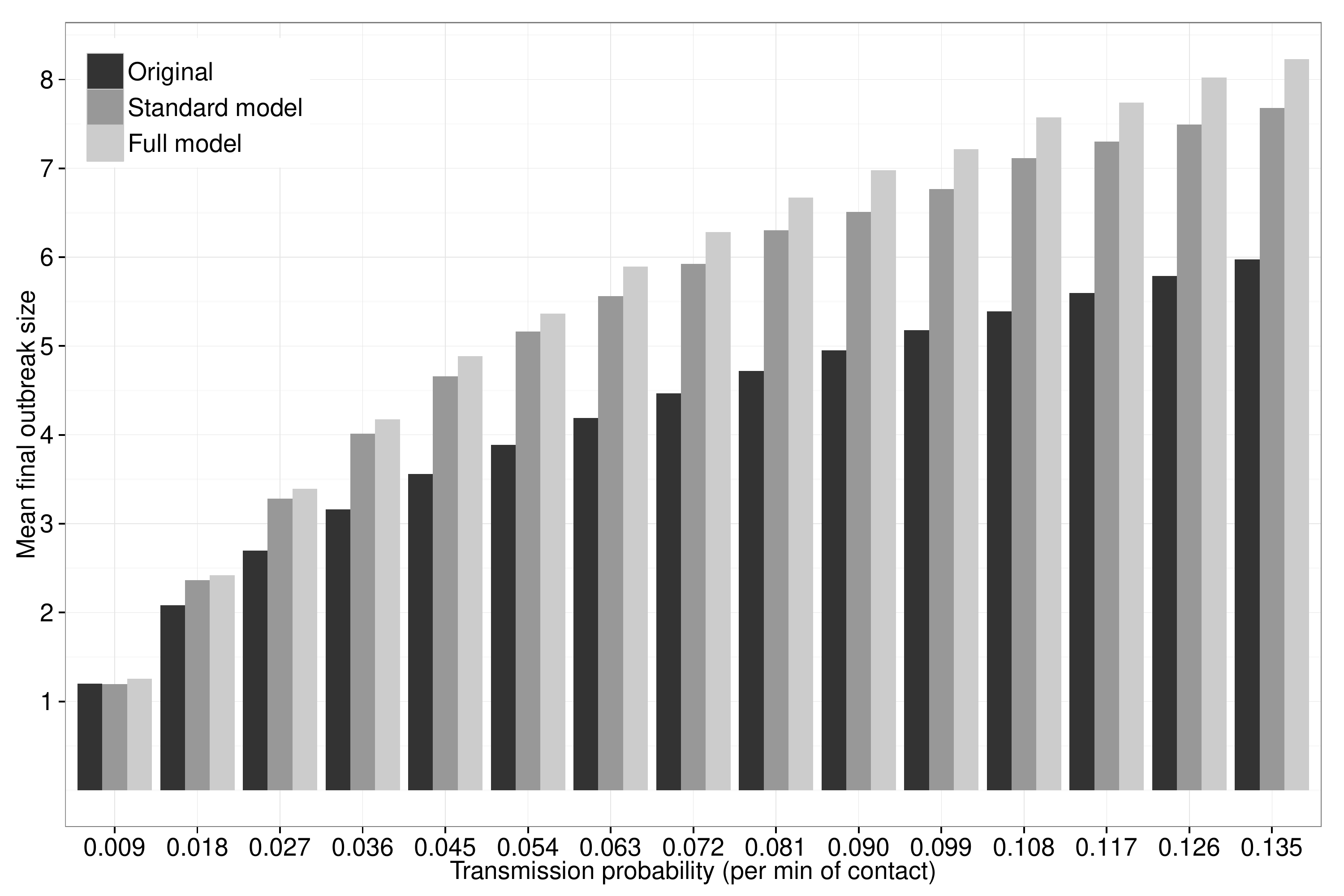}
\end{centering}
\caption{Mean final size (minus index case) by transmission probability per minute of contact and for different contact networks, based on simulations.  `Original' refers to the empirically measured network with reporting inconsistencies resolved; `Full model' and `Standard model' refer to the corresponding models parameterized in Table 6.}
\label{fig:X1}
\end{figure}

\begin{figure}
\begin{centering}
  \includegraphics[width=5.9in,height=4in]{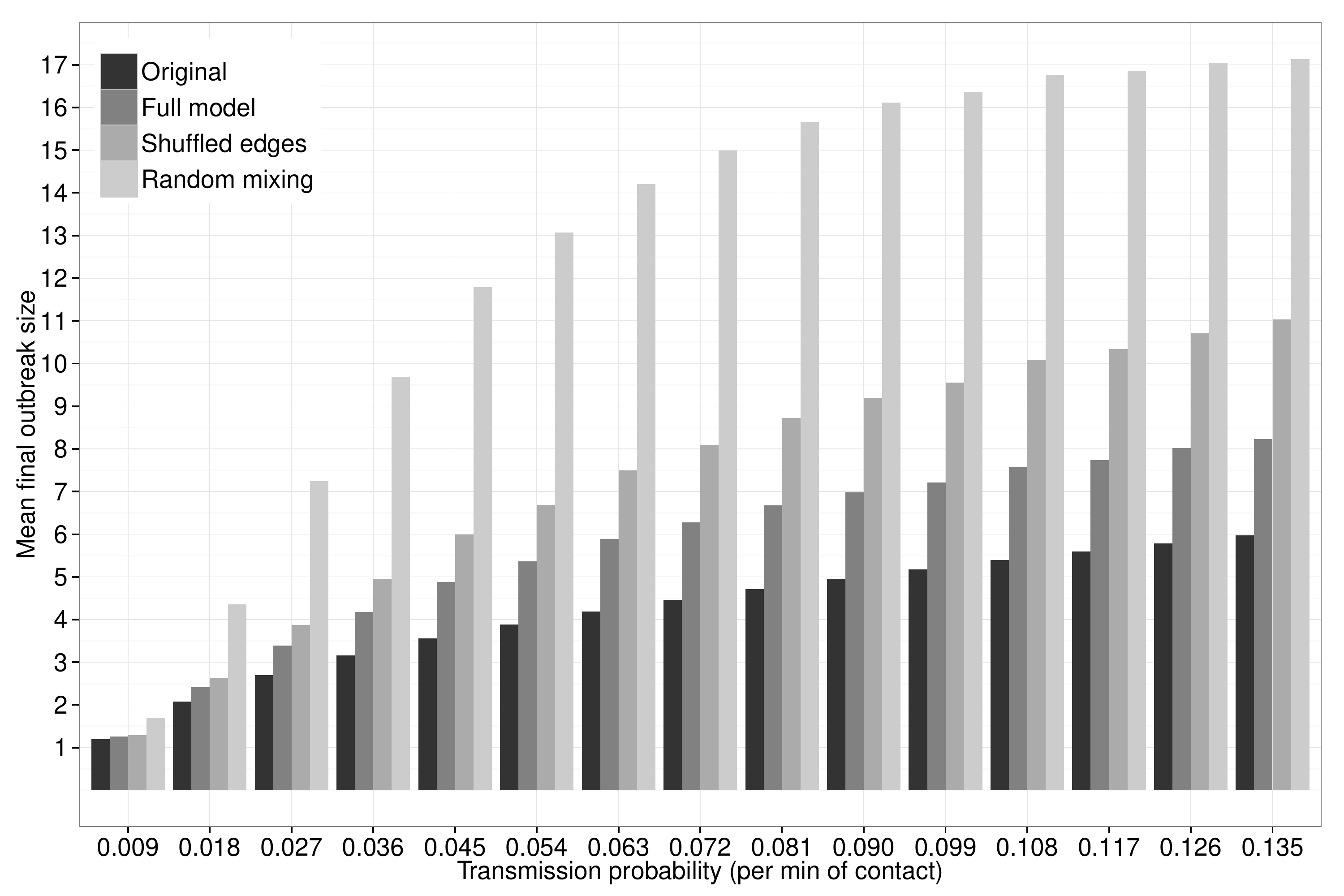}
\end{centering}
\caption{Mean final size (minus index case) by transmission probability per minute of contact and for different contact networks, based on simulations. `Original' refers to the empirically measured network with reporting inconsistencies resolved; `Full model' refers to the model in Table 6; `Shuffled edges' are network models with the same density and duration distribution as the full model, but with randomly allocated edges; `Random mixing' is a random mixing scenario.}
\label{fig:X2}
\end{figure}

\begin{figure}
\begin{centering}
  \includegraphics[width=5.9in,height=4in]{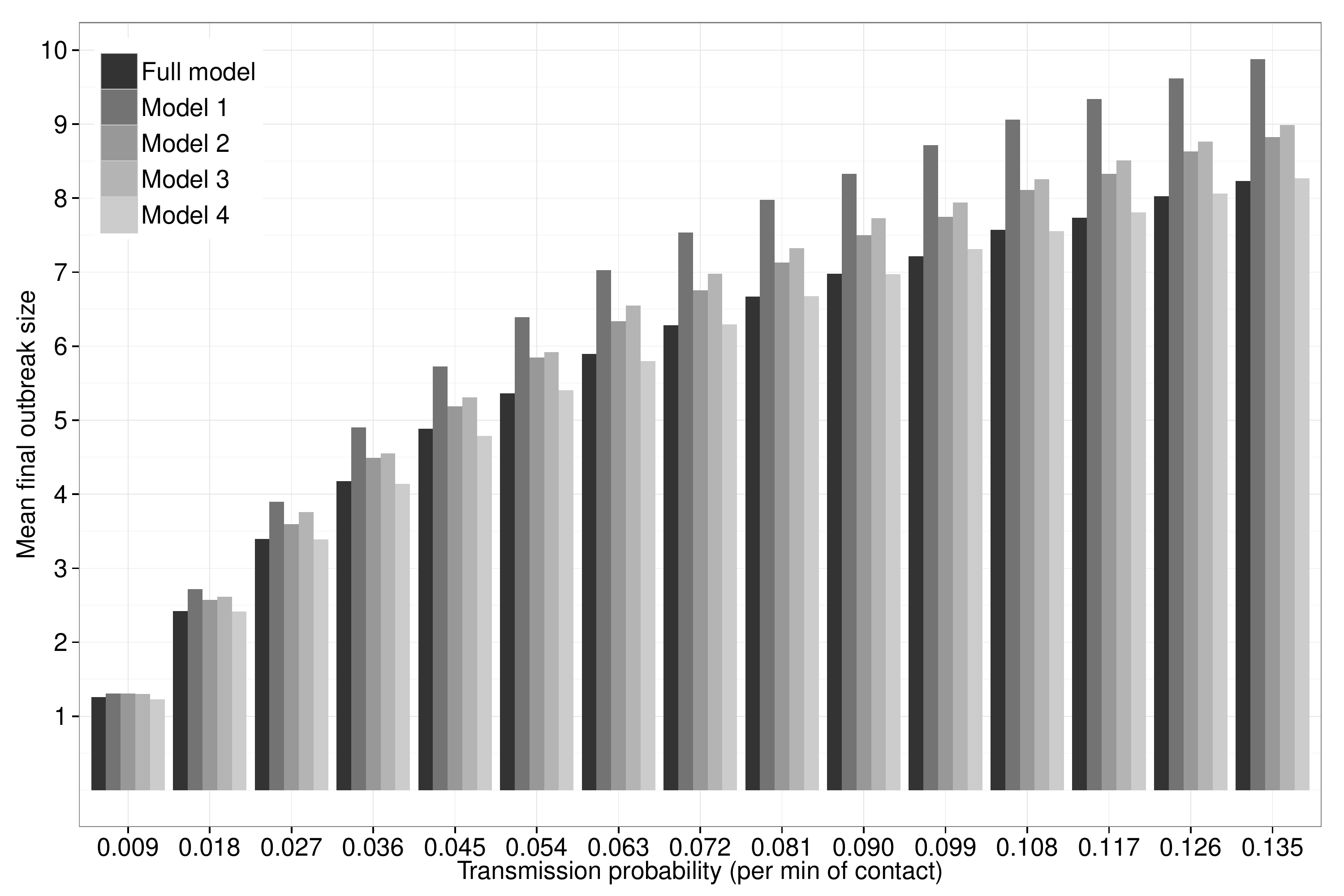}
\end{centering}
\caption{Mean final size (minus index case) by transmission probability per minute of contact for the full model and for various subsets of the full model, as defined in Table~\ref{tab:subsets}, based on simulations}
\label{fig:X3}
\end{figure}

Figures~\ref{fig:X1} --~\ref{fig:X3} show the mean final size for epidemic simulations based on various models.  We estimated 95\% confidence intervals for the mean final outbreak size using a parametric bootstrap (1000 bootstrap resamples were drawn), but these were so narrow that we omit them from the graphs.

Figure~\ref{fig:X1} compares predictions based on our model to those based on the original data.  We make this comparison as a way of assessing model fit: if we have sufficiently modelled network structure relevant to disease spread, then we expect similar predictions from the two models.  However, a shortcoming of predictions based on the empirical data is that they do not adjust for inconsistent reports, and thus, underestimate overall density.  Therefore we expect larger outbreaks from our model even if it does capture all relevant network structures.  To adjust for this, we also include our ``standard model'' (the proportional odds model not adjusting for reporting errors).  If our model captures all relevant network structure important for epidemic forecasting, the standard model should produce similar predictions to the original data.  Figure~\ref{fig:X1} shows that the adjustment for reporting error makes only a small difference in epidemic predictions, and that we have failed to capture some of the network structure important for epidemic forecasts.  However, our model's predictions are close to those based on the original data, suggesting that it does a reasonable job.

Figure~\ref{fig:X2} shows the expected final size for epidemic simulations based on the original (empirical) contact data, the full model, networks created by shuffling the edges of the full model, and a random mixing scenario.  The comparison to the shuffled edges model shows the effect of the distribution of edge duration alone, since that is the only network effect included in the shuffled edges model.  The figure shows that a large part of the difference between final size estimates based on random mixing and those based on the original data is due to the network effect of heterogeneity in edge duration.  Additional effects included in our model account for part of the additional difference, and the remaining small difference between our full model's predictions and those based on the original data remain unexplained.  All three of the network models predict smaller epidemics than a random mixing scenario, in line with what previous researchers have found (e.g., ~\cite{eames, potterschool, smieszek, szendroi, duerr}).  The clustering and repetition found in realistic contact networks tends to slow disease spread by reducing the number of susceptible persons that each infected individual comes into contact with.  

Figure~\ref{fig:X3} provides a model comparison between the full model and models 1-4 (see Table~\ref{tab:subsets}), which are subsets of the full model. While the results of all five models are very close, it is clear that the disease dynamics are closest to that of outbreaks on the original data for the full model and model 4, which differs from the full model only in its exclusion of group sociality effects.  Model 1, which only includes architectural information, performs worse than models 2 and 3, which only include organizational information.

\clearpage
\section{Discussion}
\label{section:discussion}
In this paper, we have developed social network models for face-to-face contacts relevant to infectious disease transmission in a Swiss research institute.  Our models use architectural distances between workstations, demographic variables, and organizational structure to predict contact patterns.  We found workgroup membership, collaboration on projects, mixing by employee role, and distance between workstations to be highly predictive of contact.  We found models with angular and axtopo distance measures to have higher predictive power those with metric or topo measures.  We developed latent variable models to jointly estimate duration-specific reporting probabilities and the network of contacts.  We found very high reporting probabilities for contacts with duration greater than five minutes, but only $53\%$ probability of reporting a 0--5 minute contact, consistent with findings in~\cite{smieszek2}.  We also extended our model to estimate the network of contact durations rather than the binary contact network.  Effect estimates were similar between the two models, so (for example) while collaborating on projects increases the odds of contact, it also increases the odds of longer duration contacts.  

By adjusting for measurement error and comparing to models that do not make this adjustment, we have contributed to the area of social network analysis by making a statistical improvement which can be applied to network analysis in many different settings.  Our findings have several implications for scientists.  First, our duration distribution estimate differs from that obtained by a model that does not adjust for reporting errors, since reporting probabilities vary by duration category.  Our duration distribution estimate is more accurate.  However, since reporting probabilities were very high for most durations of contacts, the difference is small, and in fact is not statistically significant.  Second, in both the binary network model and the duration network model, we found effect estimates to be only slightly (and not significantly) different from those by a model which does not adjust for reporting errors.  This finding is appealing since making this adjustment requires a more complex model, programming time, and computation time.  However, we believe that this finding is partly due to exceptionally accurate reporting and holds only under certain conditions. Other studies have found substantially less accurate reporting~\citep{read2012, test}.  In our model, we assumed the reporting probability does not depend on covariate values.  We believe that if the true reporting probability depended on covariate values, then adjusting for reporting errors would change network effect estimates.  We believe that such a condition is quite plausible.  For example, age may be related to reporting probability, as older subjects may be more likely to forget contacts.  Type of job may be as well, with those in busier and more stressful positions being more likely to forget contacts.  We explored such an extension, and our data suggest that subjects may be less likely to forget contacts occurring on different floors, but the effect was not statistically significant.  We found no evidence for a relationship between any other covariate and reporting probability.  However, our sample size is fairly small, limiting our power to detect such a relationship.  It is quite plausible that such a relationship could be detected in a larger, richer data set.  In summary, adjusting for reporting errors had only a small impact on estimates of duration distribution and network effects in this setting, but may have an impact in different settings.  We recommend further exploration into this area, including assessment of dependence between the reporting probability and covariates in the model.  In networks with lower reporting probabilities, adjustment for measurement error will be more important.  A final additional statistical improvement contributed by adjusting for reporting errors is that standard errors for network effects now include uncertainty contributed by reporting errors, so are more accurate than those in the standard model.  They are only slightly larger than those from a standard model, so the standard assumption does not do much harm.  This is because reporting probabilities are fairly high.  

Our epidemic simulation study showed that all network models considered produced epidemic forecasts closer to those based on the original data than a random mixing assumption, which tends to overestimate final size.  Predicted final sizes based on our full model were similar to those based on the original data but were not identical.  Part of this is due to the fact that our model adjusts for reporting errors, changing the estimate of the duration distribution.  The remaining difference suggests that our model omits some network structures relevant to epidemic forecasting.  We recommend further research in this area and hypothesize that additional clustering (perhaps due to friendship structure) and duration distribution may be important effects to include.  Among subsets of our model that we considered, Model 4 (which also includes architectural, demographic, and organizational data) does the best job of reproducing outbreak size, and its predictions are similar to those based on our full model.  Model 4 differs only from the full model in its exclusion of sociality effects for research groups.  Since these were not significant in the full model, we, in fact, do not have evidence that these effects are nonzero, and we consider Model 4 to be the best-fitting model.  Our analyses of Kendall's $\tau$ show that both the full model and Model 4 capture a relevant part of the individual epidemiological importance of individuals, measured as the expected number of secondary cases they would generate if infected.  While they are omitting a lot of information, the vastly reduced models 1-3 still capture some information about the individual importance of individuals. Hence, it might be worth incorporating easily accessible information about the organizational and architectural structure in future analyses and modelling efforts (see discussions in~\cite{smieszek2013low} and~\cite{chowell}).

Our work contains several limitations.  First, not everyone responded to the survey.  In particular, no professors replied, so we have a biased subset of network actors.  It is likely that the contact patterns of professors differ from those represented in our sample, but we have no information regarding how they may differ.  It is therefore best to consider our estimates to apply to the network of graduate students, postdocs, and administrative staff, rather than representing the entire contact network within the research institute.  

Next, although we adjusted for errors in reporting whether contact occurred or not, we did not adjust for inconsistencies in duration reports. Instead, we made the simplifying assumption that if two participants reported different durations for the same contact, the longer duration is the correct report. Our model could be extended to model actual rather than observed durations, for example, by including and estimating a probability that the two reported durations are within a specified threshold.  This would be an interesting avenue for future work, as there was a fair amount of inconsistency in duration reports in the data set (detailed in~\cite{smieszek2}).  We do not have information to estimate a tendency to underestimate or overestimate contact durations.  One way to collect such data might be studies that jointly collect contact diary data and wireless sensor data detecting when subjects are in close proximity and face-to-face (such as radio frequency identification data).  The first and (to our knowledge), only such study was done by~\cite{test}, who found that survey reporting was reasonably accurate for long contacts, but highly inaccurate for short contacts.  However, they found additional unexplained differences between survey and sensor data, and recommend further research into the cause of these.  Another recent paper compares online survey reports (rather than contact diaries) with wireless sensor data and direct observations, and finds major differences in network structures and interaction patterns captured by each of the methods~\citep{sailer13}.

We made the assumption that no contacts were fabricated.  We believe that false positive reports are much less likely to occur than false negatives, but integrating a probability of fabrication into our model is one possible direction for future research.  To do so, we would need an additional source of data (e.g. sensor data or observations) to judge whether inconsistent contact reports were due to fabrication or forgetting.

An additional assumption we made is the independence in contact between dyads, conditional on the effects in our model.  Further research could explore the effect of higher-order dependencies in contact patterns, such as a transitivity effect.  In the ERGM framework such an effect could be naively modelled with an inclusion of a triangles statistic, capturing the increase in the odds of contact between two people who contact the same third person, controlling for other effects in the model.  It is modelled more realistically by parametric terms which capture decreasing marginal returns on the number of mutual contacts on the increase in odds of contact~\citep{hunter2007curved}.  Including such an effect would be quite complicated in this setting.  	We did perform goodness of fit diagnostics comparing our ERGM fit to the network in which a contact was assumed to occur if reported by at least one of the members, to a model identical to ours but which adds one such transitivity term. These are included in the Appendix and indicate that our model captured a good part of transitivity present in the network, but could be improved with an additional transitivity term.  However, a previous study modelling social networks for influenza transmission in a high school found that inclusion of mixing preferences and heterogeneity in contact duration was sufficient to capture the level of clustering relevant to disease transmission, and transitivity terms did not increase predictive power of the model~\citep{potterschool}.  For this reason, we hypothesize that including network dependency terms would not improve our model for the purpose of disease prediction.  However, we suspect that we have omitted some group mixing terms or individual-level predictors from our model which may be important to explain the clustering in our network.  

In conclusion, we have used organizational structure, architectural structure, and demographic information to estimate the contact network in a research institute.  We adjusted for reporting errors and investigated the impact of this adjustment on network effect estimates.  We found the adjustment in this context to have a negligible impact on estimates, although we believe the impact may be substantial in networks in other settings.  We found angular distance, workgroup membership, and project collaboration to be highly predictive of contact in this setting.  Our epidemic simulation study shows that the network structures we have modelled are important for epidemic predictions and produce different forecasts than a random mixing assumption.  Our findings indicate that incorporating architectural and organizational data into large-scale epidemic forecasting models may improve the accuracy of epidemic predictions, thus improving our ability to contain and control epidemics.

\newpage
\bibliographystyle{nws}
\bibliography{nwsbib} 

\section{Appendices}
\appendix	
\section{Comparison of reporting probability estimates to those in previous work}

We compare the reporting probability estimates from our proportional odds model with angular distance to those from~\cite{smieszek2} in table~\ref{tab: reporting}.  The estimates obtained by the two different methods are extremely similar.  The wide confidence interval for contacts lasting more than an hour is due to the fact that all contacts of this duration were reported with 100\% consistency, so there is no variability with which to estimate the standard error of the reporting probability.

\begin{table}[ht]
\begin{center}
\caption{Comparison of our reporting probability estimates to those in Smieszek et al. (Monday only)}
\label{tab: reporting}
\setlength{\tabcolsep}{6pt}
\begin{tabular}{lrrrl}
\hline
Estimate & Angular Model& & Smieszek et al.\\ 
  \hline
0--5 & 0.56 [0.41, 0.69] && 0.53\\ 
  6--15 & 0.96 [0.84, 0.99] && 0.96\\ 
  16--60 & 0.93 [0.83, 0.99] && 0.93\\ 
  61--480 & 1.00 [0.00, 1.00] && 1.00\\ 
\hline
\end{tabular}
\end{center}
\end{table}

\newpage
\section{Results from proportional odds models with four different distance metrics}

Table~\ref{tab:POM01} shows results from proportional odds models with four different distance metrics.  The model with angular distance metrics fits best according to the AIC.

% latex table generated in R 2.13.1 by xtable 1.7-0 package
% Fri Jun 15 17:17:59 2012
\begin{table}[ht]
\begin{center}
\caption{Coefficients for proportional odds models for contact duration, using four different distance metrics}
\setlength{\tabcolsep}{6pt}
\label{tab:POM01}
\begin{tabular}{lrlrlrlrlrl}
  \hline
 & Metric & & Topo &  & Angular & & Axtopo & \\ 
%Intercepts\\
%\hline
%0 & 4.48 (1.02) & *** & 4.83 (1.10) & *** & 2.65 (1.04) & * & 2.46 (1.14) & * \\ 
 % 1-5 & 5.67 (1.03) & *** & 6.02 (1.12) & *** & 3.87 (1.04) & *** & 3.68 (1.14) & ** \\ 
 % 6-15 & 6.37 (1.04) & *** & 6.72 (1.12) & *** & 4.58 (1.04) & *** & 4.39 (1.14) & *** \\ 
 % 16-60 & 8.01 (1.07) & *** & 8.36 (1.15) & *** & 6.22 (1.08) & *** & 6.04 (1.17) & *** \\ 
%\hline
%Coefficients\\
  \hline
 Group 1 & -0.32 (0.19) & . & -0.33 (0.19) & . & -0.18 (0.20) &  & -0.20 (0.20) &  \\ 
  Group 2 & -0.07 (0.18) &  & -0.06 (0.18) &  & 0.11 (0.19) &  & 0.13 (0.20) &  \\ 
  Group mixing & 3.41 (0.48) & *** & 3.49 (0.47) & *** & 3.42 (0.45) & *** & 3.39 (0.45) & *** \\ 
  Distance & -0.01 (0.02) &  & -0.01 (0.04) &  & -0.22 (0.08) & ** & -0.20 (0.08) & * \\ 
  Female & 0.36 (0.21) & . & 0.37 (0.21) & . & 0.31 (0.21) &  & 0.31 (0.21) &  \\ 
  Role mixing & 0.79 (0.30) & ** & 0.83 (0.30) & ** & 0.60 (0.29) & * & 0.63 (0.29) & * \\ 
  Gender mixing & -0.21 (0.26) &  & -0.22 (0.26) &  & -0.18 (0.26) &  & -0.18 (0.26) &  \\ 
  Floor & 1.12 (0.52) & * & 1.23 (0.61) & * & -0.09 (0.68) &  & -0.33 (0.77) &  \\ 
  Shared projects & 1.17 (0.28) & *** & 1.20 (0.28) & *** & 1.06 (0.28) & *** & 1.08 (0.28) & *** \\ 
\hline
\\
  AIC & 779.1 &  & 779.4 &  & 772.5 &  & 773.1 &  \\ 
  \hline
\end{tabular}
\end{center}
Significance levels: *** $=p<0.001$; ** $=p<0.01$; * $=p<0.05$;  ``.'' $=p<0.10$
\end{table}
\newpage
\section{Results from testing proportional odds model assumption}

Table~\ref{tab:test POM} compares log odds ratio estimates from logistic regression models fitted to contact duration, dichotomized at different cutoffs (0, 5, 15, or 60 minutes).  Some estimates are effectively infinite, with infinite standard errors because either 0\% or 100\% cell counts were observed.  The table suggests that while the proportional odds assumption probably does not hold perfectly, it is not unreasonable.  Group mixing and distance coefficient estimates are remarkably similar, the two main effects of primary interest.  Other coefficients vary somewhat, but differences are not statistically significant.  

\begin{table}[!h]
\begin{center}
\caption{Log odds ratio estimates and 95\% confidence intervals at different dichotomizations of contact duration to test proportional odds model assumption, metric distance measure.}
\label{tab:test POM}
\begin{tabular}{lrrrrr}
  \hline
\multicolumn{1}{l}{}& \multicolumn{4}{c} {Duration cutoff }\\ 
 Effect&  $>0$& $>5$ & $>15$ & $>60$ \\ 
  \hline
%Intercept & -5.18 & -21.38 & -21.22 & -37.01 \\ 
% & [-7.59, 2.78]  & [NA, NA] & [NA, NA] & [NA, NA] \\ 
%\\
  Group 1 & -0.04 & -0.02 & -0.49 & -0.65 \\ 
 & [-0.54, 0.46] & [-0.65, 0.60] & [-1.17, 0.20] & [-1.23, -0.06] \\ 
\\
Group 2 & -0.27 & -0.05 & -0.41 & -0.58 \\ 
 & [-0.75, 0.22] & [-0.63, 0.53] & [-1.06, 0.24] & [-1.09, -0.06] \\ 
\\
Group mixing  & 3.96 & 4.13 & 3.59 & 17.73 \\ 
 & [2.92, 5.01] & [2.57, 5.69] & [2.00, 5.18] & [NA, NA]\\ 
\\
Distance  & -0.01 & -0.03 & -0.02 & -0.01 \\ 
 & [-0.04, 0.03] & [-0.07, 0.01] & [-0.06, 0.02] & [-0.06, 0.04] \\ 
\\
  Sex & -0.08 & -0.04 & 0.33 & 0.19 \\ 
 & [-0.51, 0.35] & [-0.53, 0.44] & [-0.21, 0.87] & [-0.47, 0.86] \\ 
\\
  Role & 0.93 & 1.52 & 1.76 & -0.17 \\ 
 & [0.35, 1.51] & [0.86, 2.18] & [1.01, 2.51] & [-1.28, 0.94] \\ 
\\
  Gender mixing & -0.12 & -0.29 & -0.41 & 0.7 \\ 
 & [-0.66, 0.41] & [-0.91, 0.33] & [-1.12, 0.30] & [-0.24, 1.64] \\ 
\\
  Same floor  & 1.57 & 16.82 & 16.62 & 16.93 \\ 
 & [0.40, 2.74] & [NA, NA] & [NA, NA] & [NA, NA] \\ 
\\
Shared projects & 3.78 & 2.15 & 2.39 & 1.42 \\ 
 & [1.40, 6.15]  & [0.90, 3.39] & [1.23, 3.55] & [0.32, 2.53] \\ 
  \hline
\end{tabular}
\end{center}
\end{table}

\newpage
\section{Additional fits of proportional odds models}

\begin{table}[ht]
\begin{center}
\caption{Coefficients (SEs) for proportional odds models for five days of the week, using angular distance metric}
\label{tab:POM}
\setlength{\tabcolsep}{4pt}
\begin{tabular}{lrlrlrlrlrl}
  \hline
Intercepts & Monday &  & Tuesday &  & Wednesday &  & Thursday &  & Friday &  \\ 
 \hline
0 & 2.65 (1.04) & * & 3.37 (0.88) & ** & 3.11 (1.14) & * & 4.25 (0.98) & ** & 2.98 (1.4) & * \\ 
 1-5 & 3.87 (1.04) & *** & 4.23 (0.89) & *** & 4.41 (1.13) & *** & 5.01 (0.98) & *** & 3.33 (1.40) & *** \\ 
 6-15 & 4.58 (1.04) & *** & 4.82 (0.89) & *** & 5.01 (1.13) & *** & 5.51 (0.99) & *** & 3.76 (1.40) & *** \\ 
 16-60 & 6.22 (1.08) & *** & 6.44 (0.93) & *** & 6.78 (1.15) & *** & 6.90 (1.00) & *** & 5.55 (1.41) & *** \\ 
\hline
Group 1 &	 -0.18 (0.20) &	 &	 0.40 (0.18) &	 * &	0.28 (0.18) &	 &	0.07 (0.18) &	 &	0.29 (0.31) &	 \\
Group 2 &	 0.11 (0.19) &	 &	 0.17 (0.20) &	 &	- 0.09 (0.18) &	 &	0.08 (0.19) &	 &	0.06 (0.33) &	 \\
Group mixing &	 3.42 (0.45) &	 *** &	 2.75 (0.39) &	 *** &	4.54 (0.64) &	 *** &	3.94 (0.49) &	 *** &	3.00 (0.48) &	 *** \\
Distance &	 -0.22 (0.08) &	 * &	 -0.13 (0.07) &	 . &	 -0.29 (0.09) &	 ** &	 -0.16 (0.07) &	 * &	-0.15 (0.11) &	 \\
Female &	 0.31 (0.21) &	 &	 -0.09 (0.17) &	 &	0.29 (0.20) &	 &	0.02 (0.19) &	 &	- 0.02 (0.27) &	 \\
Role mixing &	 0.60 (0.29) &	 . &	 0.29 (0.25) &	 &	0.79 (0.30) &	 * &	0.98 (0.24) &	 ** &	1.35 (0.37) &	 ** \\
Gender mixing &	 -0.18 (0.26) &	 &	 0.26 (0.22) &	 &	0.27 (0.25) &	 &	0.01 (0.22) &	 &	-0.50 (0.31) &	 \\
Floor &	 -0.09 (0.68) &	 &	 -0.14 (0.56) &	 &	-0.76 (0.69) &	 &	0.63 (0.62) &	 &	-0.61 (0.79) &	 \\
Shared projects &	 1.06 (0.28) &	 ** &	 1.62 (0.23) &	 *** &	1.28 (0.23) &	 *** &	0.82 (0.22) &	 ** &	0.65 (0.26) &	 * \\
\hline
\end{tabular}
\end{center}
\end{table}

\section{Multinomial logit model }

\subsubsection{Multinomial logit model likelihood}
\label{subsection:multinomial}

In this model we predict both contact and contact duration as a function of covariates.  We use a multinomial logit model to estimate the probability of each of the four duration categories, or a fifth category, non-contact.  We will now re-define our notation to reflect the inclusion of non-contact as a duration category.  Define $\pi_k(x) = P(D_{ij}=d_k | X_{ij}=x)$, for $k=1, \ldots, 5$ (representing categories are 0, 1-5, 6-15, 16-60, and 61+ minutes).  Let $X_{ij}$ denote individual-level and dyadic covariates in our model.  Again we let $\bD$ denote the matrix of contact durations (after removing inconsistencies in duration reports) with non-contacts having duration zero.  Using non-contact as the baseline duration category, the multinomial model is defined by \cite{agresti}:

\[
\log \frac{P(D_{ij}=d_k | X_{ij} = \mathbf{x})}{P(D_{ij}=d_1 | X_{ij} = \mathbf{x})} = \alpha_k + \beta_k^T \mathbf{x}, \textnormal{ for } k=2, 3, 4, 5
\]
From this we obtain:
\[
P(D_{ij}=d_k | X_{ij} = \mathbf{x}) = \frac{e^{\alpha_k+\beta_k^T \mathbf{x}} } {1+\sum_{h=2}^5   e^{\alpha_h+\beta_h^T \mathbf{x}}}
\]
Because the probabilities must sum to one, $P(D_{ij}=d_1 | X_{ij} = x) = 1-\sum_{h=2}^5   e^{\alpha_h+\beta_h^T \mathbf{x}}$.

By applying our assumptions, rules of conditional probability, and the Law of Total Probability, we find that the joint likelihood of $\bD$ and $\bC$ is:
\[
P(C_{ij}=1, C_{ji}=1, D_{ij}=d_k)= P(D_{ij}=d_k)p_k^2
\]
\[
P(C_{ij}=1, C_{ji}=0, D_{ij}=d_k)=P(D_{ij}=d_k) p_k(1-p_k)
\]
\[
P(C_{ij}=0, C_{ji}=0, D_{ij}=0)= P(D_{ij}=0) + \sum_{k=2}^5 P(D_{ij}=d_k) (1-p_k)^2 
\]

Then the probability of the observed data is:
\[
P(\bC=\bc, \bD=\bd) = \prod_{i=1}^n \prod_{j=i+1}^n P(C_{ij}=c_{ij}, C_{ji}=c_{ji}, D_{ij}=d_k)
\]
We maximize the log likelihood to estimate $\balpha$, $\bbeta$, and $\bf{p}$ using the {\tt trust} function in R and computed standard errors by inverting the Fisher information matrix~\citep{trust}.

\newpage

\section{Goodness of fit to assess modelling of transitivity}

\begin{figure}[htbp]
\begin{centering}
  \includegraphics[width=5.5in,height=4.5in]{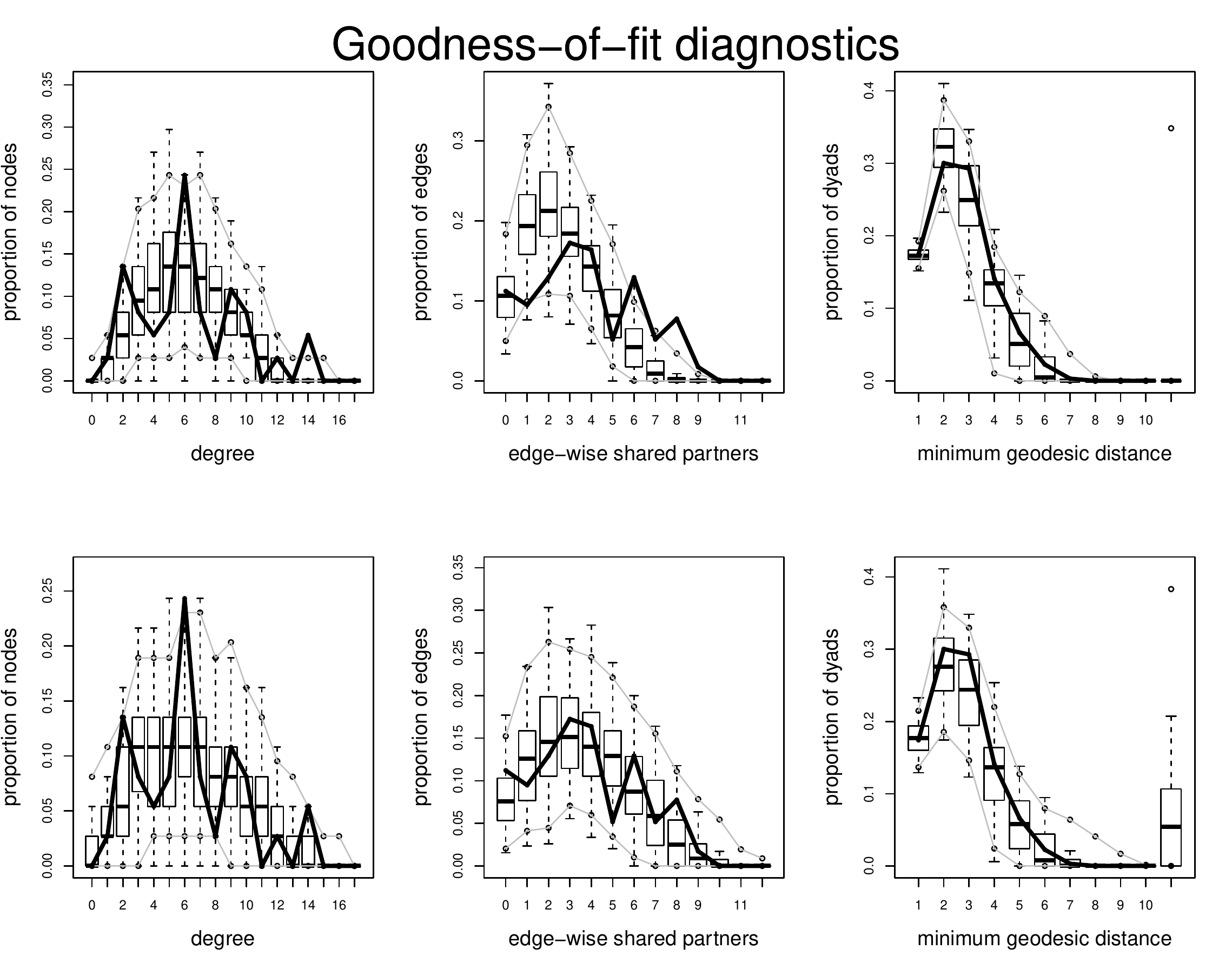}
\end{centering}
\caption{\label{fig:gof}  Goodness of fit diagnostics for our model (top) without adjusting for reporting errors, compared to those for an extension of model which also includes a gwesp(0.5) term to capture transitivity.}
\end{figure}

Figure~\ref{fig:gof} compares goodness of fit diagnostics for two models in order to assess how well our model captured transitivity present in the network.  The first model is our ERGM with angular distance, fit to a nondirectional binary network created by assuming that contact between two individuals occurred if it was reported by at least one of the two.  The second model is the same ERGM, but also including a geometrically weighted edgewise shared partners (gwesp) term with alpha = 0.5.  The box plots show network statistics for networks simulated from each model, while the solid line shows network statistics for the actual data.  The figures show that our model does a good job representing the degree distribution and the minimum geodesic distance of the network, but overestimates the proportion of edges with 2--3 shared partners, and underestimates the proportion of edges with 6--8 shared partners.  The model with the added gwesp term mostly corrects this problem.

\newpage

\subsection{Multinomial logit model likelihood results}

Table~\ref{tab:multinomial1} shows coefficient estimates from the multinomial logit model with four distance metrics.  Coefficients are interpreted as follows: The odds of a 1--5 minute contact relative to no contact increases by a factor of $e^{3.24}=26$ if two people are in the same research group, controlling for other variables in the model.  The odds of a 16--60 minute contact relative to no contact decreases by a factor of $e^{-0.05}=0.95$ for each unit increase in metric distance between their workstations, controlling for other variables in the model.  Some coefficients do not have finite standard errors because of zero or 100\% cell counts.  For example, all reported 16--60 and 60+ minute contacts were on the same floor.  The floor coefficient for these categories should be infinite, but is estimated as a very large number (after exponentiation).  All reported 61+ minute contacts were among members of the same research group, resulting in an infinite coefficient for group mixing.  The set of predictor	 variables in the multinomial model that we fit differs from our full model in the text in that the shared projects is excluded.  However, inclusion of this variable would only amplify the estimation problems caused by a large number of parameters being estimated with several cases of small cell counts. 

We include in this section estimates from the proportional odds model so the reader may compare them to the multinomial model.  

 %The intercept terms counterbalances these infinite coefficient estimates.

\begin{table}[ht]
\begin{center}
\caption{Multinomial model estimates (SEs)}
\footnotesize
\label{tab:multinomial1}
\begin{tabular}{lrlrlrlrlrl}
  \hline
 & Metric & & Angular & & Topo& & Axtopo & \\ 
  \hline
1--5 minutes\\
\hline
Int. & -4.48 (0.94) & *** & -2.28 (1.05) & * & -4.85 (1.02) & *** & -2.36 (1.18) & * \\ 
  Group 1 & -0.01 (0.2) &  & 0.15 (0.21) &  & -0.02 (0.2) &  & 0.11 (0.21) &  \\ 
  Group 2 & 0.29 (0.2) &  & 0.5 (0.22) & * & 0.28 (0.2) &  & 0.49 (0.23) & * \\ 
  Group Mixing & 3.24 (0.44) & *** & 3.22 (0.41) & *** & 3.28 (0.43) & *** & 3.19 (0.41) & *** \\ 
  Distance & 0 (0.02) &  & -0.21 (0.1) & * & 0.02 (0.04) &  & -0.16 (0.09) & . \\ 
  Female & -0.13 (0.21) &  & -0.15 (0.21) &  & -0.13 (0.21) &  & -0.15 (0.21) &  \\ 
  Role Mixing & 0.42 (0.31) &  & 0.24 (0.32) &  & 0.45 (0.31) &  & 0.29 (0.31) &  \\ 
  Gender Mixing & -0.3 (0.26) &  & -0.28 (0.26) &  & -0.3 (0.26) &  & -0.29 (0.26) &  \\ 
  Floor & 0.22 (0.48) &  & -1.15 (0.72) &  & 0.42 (0.57) &  & -1.2 (0.84) &  \\ 
 \hline
6--15 minutes\\
\hline
Int.  & -6.6 (1.47) & *** & -2.34 (1.42) &  & -6.6 (1.58) & *** & -1.91 (1.53) &  \\ 
  Group 1  & -0.08 (0.26) &  & 0.31 (0.28) &  & -0.08 (0.26) &  & 0.25 (0.27) &  \\ 
  Group 2  & 0.37 (0.25) &  & 0.86 (0.28) & ** & 0.4 (0.26) &  & 0.89 (0.28) & ** \\ 
  Group Mixing  & 3.63 (0.79) & *** & 3.78 (0.78) & *** & 3.68 (0.79) & *** & 3.7 (0.78) & *** \\ 
  Distance  & -0.02 (0.02) &  & -0.55 (0.11) & *** & -0.05 (0.05) &  & -0.48 (0.1) & *** \\ 
  Female  & 0.21 (0.26) &  & 0.09 (0.27) &  & 0.21 (0.26) &  & 0.09 (0.26) &  \\ 
  Role Mixing  & 0.9 (0.37) & * & 0.48 (0.37) &  & 0.91 (0.37) & * & 0.53 (0.37) &  \\ 
  Gender Mixing  & -0.23 (0.33) &  & -0.17 (0.34) &  & -0.24 (0.33) &  & -0.17 (0.34) &  \\ 
  Floor  & 1.21 (0.88) &  & -1.84 (1.06) & . & 1.05 (0.98) &  & -2.4 (1.17) & * \\ 
  \hline
16--60 minutes\\
\hline
Int. & -19.76 (NA) &  & -18.94 (NA) &  & -19.94 (NA) &  & -18.23 (NA) &  \\ 
  Group 1 & -0.12 (0.23) &  & 0.09 (0.24) &  & -0.13 (0.23) &  & 0.05 (0.23) &  \\ 
  Group 2 & -0.09 (0.23) &  & 0.26 (0.24) &  & -0.04 (0.23) &  & 0.28 (0.24) &  \\ 
  Group Mixing & 3.72 (0.78) & *** & 4.05 (0.75) & *** & 3.75 (0.78) & *** & 4 (0.75) & *** \\ 
  Distance & -0.05 (0.02) & ** & -0.39 (0.1) & *** & -0.1 (0.04) & * & -0.35 (0.1) & *** \\ 
  Female & 1.1 (0.4) & ** & 1.05 (0.41) & * & 1.12 (0.4) & ** & 1.05 (0.41) & * \\ 
  Role Mixing & 1.58 (0.38) & *** & 1.53 (0.37) & *** & 1.6 (0.38) & *** & 1.56 (0.37) & *** \\ 
  Gender Mixing & -1.31 (0.45) & ** & -1.38 (0.45) & ** & -1.33 (0.45) & ** & -1.37 (0.45) & ** \\ 
  Floor & 14.49 (NA) &  & 13.8 (NA) &  & 14.5 (NA) &  & 12.99 (NA) &  \\ 
  \hline
61+ minutes\\
\hline
Int.  & -26.51 (NA) &  & -52.12 (6.45) & *** & -29.14 (NA) &  & -24.45 (NA) &  \\ 
  Group 1  & -0.74 (0.3) & * & -0.5 (0.31) &  & -0.72 (0.3) & * & -0.53 (0.31) & . \\ 
  Group 2  & -0.61 (0.26) & * & -0.27 (0.28) &  & -0.56 (0.26) & * & -0.21 (0.29) &  \\ 
  Group Mixing  & 13.72 (126.44) &  & 42.44 (10.25) & *** & 14.7 (116.34) &  & 14.09 (NA) &  \\ 
  Distance  & -0.04 (0.03) &  & -0.35 (0.15) & * & -0.03 (0.06) &  & -0.34 (0.14) & * \\ 
  Female  & 0.38 (0.34) &  & 0.3 (0.34) &  & 0.35 (0.34) &  & 0.31 (0.34) &  \\ 
  Role Mixing  & 0.31 (0.54) &  & 0.28 (0.51) &  & 0.53 (0.54) &  & 0.27 (0.52) &  \\ 
  Gender Mixing  & 0.53 (0.48) &  & 0.47 (0.48) &  & 0.47 (0.48) &  & 0.48 (0.48) &  \\ 
  Floor  & 11.83 (NA) &  & 9.15 (NA) &  & 12.86 (NA) &  & 9.78 (NA) &  \\ 
\hline
\\
AIC   & 1478 &  & 1453 &  & 1480 &  & 1456 &  \\ 
   \hline
\end{tabular}
\end{center}
Significance levels: *** $=p<0.001$; ** $=p<0.01$; * $=p<0.05$;  ``.'' $=p<0.10$
\end{table}

% latex table generated in R 2.13.1 by xtable 1.7-0 package
% Fri Jun 15 17:17:59 2012
\begin{table}[ht]
\begin{center}
\caption{Coefficients (SEs) for proportional odds models for contact duration, using four different distance metrics}
\setlength{\tabcolsep}{6pt}
\label{tab:POM1}
\begin{tabular}{lrlrlrlrlrl}
  \hline
 %& Metric & & Topo &  & Angular & & Axtopo & \\ 
%Intercepts\\
%\hline
%0 & 4.48 (1.02) & *** & 4.83 (1.10) & *** & 2.65 (1.04) & * & 2.46 (1.14) & * \\ 
 % 1-5 & 5.67 (1.03) & *** & 6.02 (1.12) & *** & 3.87 (1.04) & *** & 3.68 (1.14) & ** \\ 
 % 6-15 & 6.37 (1.04) & *** & 6.72 (1.12) & *** & 4.58 (1.04) & *** & 4.39 (1.14) & *** \\ 
 % 16-60 & 8.01 (1.07) & *** & 8.36 (1.15) & *** & 6.22 (1.08) & *** & 6.04 (1.17) & *** \\ 
%\hline
%Coefficients\\
%  \hline
 %Group 1 & -0.32 (0.19) & . & -0.33 (0.19) & . & -0.18 (0.20) &  & -0.20 (0.20) &  \\ 
  %Group 2 & -0.07 (0.18) &  & -0.06 (0.18) &  & 0.11 (0.19) &  & 0.13 (0.20) &  \\ 
 % Group mixing & 3.41 (0.48) & *** & 3.49 (0.47) & *** & 3.42 (0.45) & *** & 3.39 (0.45) & *** \\ 
 % Distance & -0.01 (0.02) &  & -0.01 (0.04) &  & -0.22 (0.08) & ** & -0.20 (0.08) & * \\ 
  %Female & 0.36 (0.21) & . & 0.37 (0.21) & . & 0.31 (0.21) &  & 0.31 (0.21) &  \\ 
 % Role mixing & 0.79 (0.30) & ** & 0.83 (0.30) & ** & 0.60 (0.29) & * & 0.63 (0.29) & * \\ 
 % Gender mixing & -0.21 (0.26) &  & -0.22 (0.26) &  & -0.18 (0.26) &  & -0.18 (0.26) &  \\ 
  %Floor & 1.12 (0.52) & * & 1.23 (0.61) & * & -0.09 (0.68) &  & -0.33 (0.77) &  \\ 
  %Shared projects & 1.17 (0.28) & *** & 1.20 (0.28) & *** & 1.06 (0.28) & *** & 1.08 (0.28) & *** \\ 
&	 Metric &	 &	 Angular &	 &	 Topo &	 &	 Axtopo &	 \\
%Intercepts\\								
%\hline								
%0 &	 4.48 (1.02) &	 *** &	 2.65 (1.04) &	 * &	 4.83 (1.10) &	 *** &	 2.46 (1.14) &	 * \\
% 1-5 &	 5.67 (1.03) &	 *** &	 3.87 (1.04) &	 *** &	 6.02 (1.12) &	 *** &	 3.68 (1.14) &	 ** \\
% 6-15 &	 6.37 (1.04) &	 *** &	 4.58 (1.04) &	 *** &	 6.72 (1.12) &	 *** &	 4.39 (1.14) &	 *** \\
% 16-60 &	 8.01 (1.07) &	 *** &	 6.22 (1.08) &	 *** &	 8.36 (1.15) &	 *** &	 6.04 (1.17) &	 *** \\
%\hline								
%Coefficients\\								
\hline
Group 1 &	 -0.32 (0.19) &	 . &	 -0.18 (0.20) &	 &	 -0.33 (0.19) &	 . &	 -0.20 (0.20) &	 \\
Group 2 &	 -0.07 (0.18) &	 &	 0.11 (0.19) &	 &	 -0.06 (0.18) &	 &	 0.13 (0.20) &	 \\
Group mixing &	 3.41 (0.48) &	 *** &	 3.42 (0.45) &	 *** &	 3.49 (0.47) &	 *** &	 3.39 (0.45) &	 *** \\
Distance &	 -0.01 (0.02) &	 &	 -0.22 (0.08) &	 ** &	 -0.01 (0.04) &	 &	 -0.20 (0.08) &	 * \\
Female &	 0.36 (0.21) &	 . &	 0.31 (0.21) &	 &	 0.37 (0.21) &	 . &	 0.31 (0.21) &	 \\
Role mixing &	 0.79 (0.30) &	 ** &	 0.60 (0.29) &	 * &	 0.83 (0.30) &	 ** &	 0.63 (0.29) &	 * \\
Gender mixing &	 -0.21 (0.26) &	 &	 -0.18 (0.26) &	 &	 -0.22 (0.26) &	 &	 -0.18 (0.26) &	 \\
Floor &	 1.12 (0.52) &	 * &	 -0.09 (0.68) &	 &	 1.23 (0.61) &	 * &	 -0.33 (0.77) &	 \\
Shared projects &	 1.17 (0.28) &	 *** &	 1.06 (0.28) &	 *** &	 1.20 (0.28) &	 *** &	 1.08 (0.28) &	 *** \\
\hline
\\
  AIC & 779.1 &  & 772.5 &  &  779.4&  & 773.1 &  \\ 
  \hline
\end{tabular}
\end{center}
Significance levels: *** $=p<0.001$; ** $=p<0.01$; * $=p<0.05$;  ``.'' $=p<0.10$
\end{table}

\begin{table}[ht]
\begin{center}
\footnotesize
\setlength{\tabcolsep}{6pt}
\caption{Coefficients [95\% Confidence Intervals] for multinomial model with no floor effect and two largest duration categories collapsed}
\begin{tabular}{lrlrlrlrl}
  \hline
METRIC MODEL&\multicolumn{2}{c}{1-5 mins}& \multicolumn{2}{c}{6-15 mins} & \multicolumn{2}{c}{16+ mins} \\ 
Effect & Est. & 95\% CI & Est. & 95\% CI & Est. & 95\% CI \\ 
  \hline
Intercept & -4.16 & [-5.66, -2.66] & -5.22 & [-7.47, -2.97] & -4.07 & [-6.06, -2.07] \\ 
 Group 1 & -0.03 & [-0.42, 0.37] & -0.13 & [-0.66, 0.41] & -0.44 & [-0.84, -0.05] \\ 
 Group 2  & 0.31 & [-0.06, 0.69] & 0.43 & [-0.09, 0.95] & -0.30 & [-0.68, 0.08] \\ 
 Group Membership  & 3.23 & [2.36, 4.11] & 3.68 & [2.09, 5.28] & 4.14 & [2.61, 5.67] \\ 
Distance  & 0 & [-0.03, 0.02] & -0.04 & [-0.08, -0.01] & -0.07 & [-0.10, -0.04] \\ 
Sex   & -0.12 & [-0.53, 0.29] & 0.24 & [-0.27, 0.75] & 0.68 & [0.18, 1.19] \\ 
Role mixing  & 0.43 & [-0.19, 1.06] & 0.86 & [0.13, 1.59] & 1.13 & [0.49, 1.76] \\ 
Sex mixing  & -0.29 & [-0.81, 0.23] & -0.16 & [-0.82, 0.49] & -0.49 & [-1.10, 0.12] \\ 
   \hline
TOPO MODEL&\multicolumn{2}{c}{1-5 mins}& \multicolumn{2}{c}{6-15 mins} & \multicolumn{2}{c}{16+ mins} \\ 
Effect & Est. & 95\% CI & Est. & 95\% CI & Est. & 95\% CI \\ 
  \hline
Int. & -4.27 & [-5.7, -2.84] & -5.27 & [-7.47, -3.07] & -4.26 & [-6.24, -2.28] \\ 
  Group 1 & -0.03 & [-0.43, 0.37] & -0.10 & [-0.63, 0.43] & -0.42 & [-0.81, -0.03] \\ 
  Group 2 & 0.32 & [-0.06, 0.70] & 0.44 & [-0.08, 0.96] & -0.26 & [-0.64, 0.12] \\ 
  Group Mixing & 3.28 & [2.43, 4.13] & 3.68 & [2.10, 5.26] & 4.18 & [2.65, 5.71] \\ 
  Distance & 0 & [-0.05, 0.05] & -0.09 & [-0.16, -0.02] & -0.14 & [-0.20, -0.08] \\ 
  Female & -0.12 & [-0.53, 0.29] & 0.24 & [-0.27, 0.75] & 0.69 & [0.19, 1.19] \\ 
  Role Mixing & 0.43 & [-0.19, 1.05] & 0.85 & [0.12, 1.58] & 1.13 & [0.49, 1.77] \\ 
  Gender Mixing & -0.29 & [-0.81, 0.23] & -0.19 & [-0.85, 0.47] & -0.53 & [-1.14, 0.08] \\ 
   \hline
ANGULAR MODEL&\multicolumn{2}{c}{1-5 mins}& \multicolumn{2}{c}{6-15 mins} & \multicolumn{2}{c}{16+ mins} \\ 
Effect & Est. & 95\% CI & Est. & 95\% CI & Est. & 95\% CI \\ 
  \hline
Int. & -3.63 & [-4.90, -2.36] & -4.11 & [-6.13, -2.09] & -4.15 & [-6.01, -2.29] \\ 
  Group 1 & 0.09 & [-0.33, 0.51] & 0.26 & [-0.28, 0.8] & -0.13 & [-0.53, 0.27] \\ 
  Group 2 & 0.31 & [-0.06, 0.68] & 0.70 & [0.18, 1.22] & 0.01 & [-0.36, 0.38] \\ 
  Group Mixing & 3.06 & [2.28, 3.84] & 3.53 & [2.03, 5.03] & 4.37 & [2.89, 5.85] \\ 
  Distance & -0.08 & [-0.18, 0.02] & -0.45 & [-0.63, -0.27] & -0.40 & [-0.54, -0.26] \\ 
  Female & -0.16 & [-0.57, 0.25] & 0.09 & [-0.44, 0.62] & 0.57 & [0.07, 1.07] \\ 
  Role Mixing & 0.30 & [-0.31, 0.91] & 0.53 & [-0.20, 1.26] & 1.09 & [0.47, 1.71] \\ 
  Gender Mixing & -0.28 & [-0.80, 0.24] & -0.18 & [-0.85, 0.49] & -0.59 & [-1.20, 0.02] \\ 
   \hline
AXTOPO MODEL&\multicolumn{2}{c}{1-5 mins}& \multicolumn{2}{c}{6-15 mins} & \multicolumn{2}{c}{16+ mins} \\ 
Effect & Est. & 95\% CI & Est. & 95\% CI & Est. & 95\% CI \\ 
  \hline
Int. & -3.79 & [-5.05, -2.53] & -4.32 & [-6.32, -2.32] & -4.19 & [-6.04, -2.34] \\ 
  Group 1 & 0.06 & [-0.35, 0.47] & 0.18 & [-0.35, 0.71] & -0.17 & [-0.57, 0.23] \\ 
  Group 2 & 0.31 & [-0.06, 0.68] & 0.69 & [0.18, 1.20] & 0.03 & [-0.35, 0.41] \\ 
  Group Mixing & 3.09 & [2.30, 3.88] & 3.46 & [1.96, 4.96] & 4.24 & [2.76, 5.72] \\ 
  Distance & -0.05 & [-0.13, 0.03] & -0.36 & [-0.52, -0.2] & -0.35 & [-0.48, -0.22] \\ 
  Female & -0.15 & [-0.56, 0.26] & 0.10 & [-0.42, 0.62] & 0.57 & [0.07, 1.07] \\ 
  Role Mixing & 0.34 & [-0.27, 0.95] & 0.60 & [-0.13, 1.33] & 1.11 & [0.49, 1.73] \\ 
  Gender Mixing & -0.29 & [-0.81, 0.23] & -0.18 & [-0.85, 0.49] & -0.59 & [-1.2, 0.02] \\ 
\end{tabular}
\end{center}
\end{table}

\label{lastpage}

\end{document}